\documentclass[conference]{IEEEtran}

\usepackage{balance}

\IEEEoverridecommandlockouts 
\usepackage[dvips]{graphicx}
\usepackage{listings}
\usepackage[OT4,T1]{fontenc}
\usepackage[cmex10]{amsmath}
\usepackage{url}
\usepackage{multirow}
\usepackage{mathtools}
\usepackage{stmaryrd}
\usepackage{amssymb}
\usepackage{moreverb}
\usepackage{amssymb}
\usepackage{lineno}
\usepackage{graphicx}
\usepackage{amsmath}
\usepackage{moreverb}
\usepackage{amssymb}
\usepackage{lineno}
\usepackage{graphicx}
\graphicspath{ {images/} }
\usepackage{caption}
\usepackage{subcaption}
\usepackage{multirow}
\usepackage[utf8]{inputenc}
\usepackage[english]{babel}
\usepackage{mathtools}
\usepackage{babel}
\usepackage[linesnumbered,boxed,lined, ruled]{algorithm2e}
\let\oldnl\nl
\newcommand{\nonl}{\renewcommand{\nl}{\let\nl\oldnl}}
\usepackage{stackengine}
\usepackage{rotating,booktabs,comment,array,tabu,pdflscape}
\usepackage{longtable,slashbox,color,colortbl}
\usepackage{float}
\usepackage{stmaryrd}
\usepackage{multirow}
\usepackage{epstopdf}
\usepackage{makecell}
\usepackage{multirow}
\usepackage{booktabs}
\usepackage{graphicx}
\usepackage[colorlinks]{hyperref}
\usepackage{ragged2e}
\usepackage{lipsum}
\usepackage{setspace}

\title{A Distributed Application Placement and Migration Management Techniques for Edge and Fog Computing Environments}
\author{\IEEEauthorblockN{Mohammad Goudarzi\IEEEauthorrefmark{1},
Marimuthu Palaniswami\IEEEauthorrefmark{2},
Rajkumar Buyya\IEEEauthorrefmark{1}, 
\IEEEauthorblockA{\IEEEauthorrefmark{1}The Cloud Computing and Distributed Systems (CLOUDS) Laboratory \\School of Computing and Information Systems \\ The University of Melbourne, Australia\\ Email: mgoudarzi@student.unimelb.edu.au, rbuyya@unimelb.edu.au}
\IEEEauthorblockA{\IEEEauthorrefmark{2} The Department of Electrical and Electronic Engineering,\\ The University of Melbourne, Australia\\
Email: palani@unimelb.edu.au}}}

\begin{document}
\maketitle              

\begin{abstract}
Fog/Edge computing model allows harnessing of resources in the proximity of the Internet of Things (IoT) devices to support various types of latency-sensitive IoT applications. However, due to the mobility of users and a wide range of IoT applications with different resource requirements, it is a challenging issue to satisfy these applications' requirements. The execution of IoT applications exclusively on one fog/edge server may not be always feasible due to limited resources, while the execution of IoT applications on different servers requires further collaboration and management among servers. Moreover, considering user mobility, some modules of each IoT application may require migration to other servers for execution, leading to service interruption and extra execution costs. In this article, we propose a new weighted cost model for hierarchical fog computing environments, in terms of the response time of IoT applications and energy consumption of IoT devices, to minimize the cost of running IoT applications and potential migrations. Besides, a distributed clustering technique is proposed to enable the collaborative execution of tasks, emitted from application modules, among servers. Also, we propose an application placement technique to minimize the overall cost of executing IoT applications on multiple servers in a distributed manner. Furthermore, a distributed migration management technique is proposed for the potential migration of applications' modules to other remote servers as the users move along their path. Besides, failure recovery methods are embedded in the clustering, application placement, and migration management techniques to recover from unpredicted failures. The performance results demonstrate that our technique significantly improves its counterparts in terms of placement deployment time, average execution cost of tasks, the total number of migrations, the total number of interrupted tasks, and cumulative migration cost.
\end{abstract}

\section{Introduction}
The number of latency-sensitive applications of Internet of Things (IoT) devices has been increasing due to recent advances in technologies so that many applications rely on remote resources for their execution. Due to latency-sensitive nature of these applications and the huge amount of data that they generate, traditional cloud computing cannot efficiently satisfy the requirements of IoT applications, and they experience high latency and energy consumption while communicating to cloud servers (CSs) \cite{mahmud2018latency,goudarzi2020application}. The fog/edge computing paradigm addresses these issues by providing an intermediate layer of distributed resources between IoT devices and CSs that can be accessed with lower latency \cite{goudarzi2019fog,guo2018efficient,chen2015efficient}. However, the provided resources of fog/edge servers for IoT applications are limited and with less variety in comparison to the resources of CSs \cite{shckhar2019urmila}. In our view, fog computing has a hierarchical and distributed structure that harnesses the resources of both CSs and Fog Servers (FSs) at different hierarchical fog levels, while lower-level FSs have fewer computing resources compared to higher-level FSs, but they are accessible with lower latency \cite{mahmud2018latency,taneja2017resource,kiani2019hierarchical,pallewatta2019microservices,gupta2017ifogsim}. However, edge computing does not have this hierarchical structure and does not use resources of CSs \cite{yang2018distributed} (although some works use these terms interchangeably).
\par
Real-time IoT applications can be modeled as a set of lightweight and interdependent application modules in fog computing environments so that such application modules alongside their allocated resources form the data processing elements of various IoT applications \cite{taneja2017resource,mahmud2018latency}. Considering different requirements of applications' modules, they can be placed on one FS, different FSs in the same hierarchical level, FSs in different hierarchical levels, and/or CSs for the execution \cite{mahmud2018latency,jovsilo2020computation}. Besides, as the number of IoT applications increases, more requests are forwarded to FSs that may overload them. Hence, a dynamic application placement technique is required to efficiently place interdependent modules of IoT applications on remote servers while meeting their requirements.
\par    
Alongside the importance of suitable application placement techniques, there are yet several issues to be addressed. The coverage ranges of lower-level FSs are limited, and IoT users have different mobility patterns. Besides, interdependent modules of each IoT application may be deployed on several FSs. Hence, as the IoT user moves towards its destination, the application response time and IoT device energy consumption can be negatively affected \cite{wang2019delay}. Therefore, the migration of interdependent modules of each application among FSs, which incurs service interruption and additional cost, is an important and yet a challenging issue. Several migration techniques decide when, how, and where application modules can migrate when IoT users change their location in the fog/edge computing environments, such as \cite{deng2021fogbus2,machen2017live,wang2016dynamic,ouyang2018follow}. However, these techniques either focus on the migration of a single application module without considering other deployed modules \cite{wang2019delay} or consider an IoT application as a set of independent application modules. An IoT application may consist of several interdependent modules, and the migration technique should consider the configuration of all interdependent modules when an IoT user moves towards its destination. Hence, the migration of IoT applications, consisting of several interdependent modules, is an important challenge to be addressed, especially in hierarchical fog computing environments in which modules may be placed on different hierarchical levels.
\par
Also, in fog computing, there are several studies that consider the application placement and migration management engines (i.e., decision engines) have a global view about topology and resources of all FSs and CSs \cite{shckhar2019urmila,adhikari2019application} while there are other studies that assume decision engines only have a local view about resources and topology of servers in their proximity \cite{gupta2017ifogsim,pallewatta2019microservices, bittencourt2017mobility}. In these latter techniques, the decision engines act in the distributed manner so that each FS that receives the application placement and/or migration request try to use the available resources in its proximity (which can be accessed with lower latency) to place/migrate the application modules as much as possible. However, if there are no available resources, the rest of the placement and migration will be handled by higher-level FSs in the hierarchy. Considering communication with higher-level FSs incurs higher latency compared to communication among FSs at the same hierarchical level, the clustering of FSs (if it is possible) at the same hierarchical level can provide sufficient resources (with less latency in comparison to higher-level FSs) to serve real-time IoT applications and reduce the amount of communication with higher-level FSs.
\par			
In this paper, we address these issues and propose efficient distributed application placement and migration management techniques to satisfy the requirements of real-time IoT applications while users move. 
\par
The main contributions of this paper are as follows.			
\begin{itemize}
	\item We propose a new weighted cost model based on IoT applications' response time and IoT devices' energy consumption for application placement and migration of IoT devices in hierarchical fog/edge computing environments to minimize cost of running real-time IoT applications.
	
	\item We put forward a dynamic and distributed clustering technique to form clusters of FSs at the same hierarchical levels so that such servers can collaboratively handle IoT application requirements with less execution cost.
	
	\item Considering the NP-Complete nature of application placement and migration problems in fog/edge computing environments, we propose a distributed application placement and migration management techniques to place/migrate modules of real-time applications on different levels of hierarchical architecture based on their requirements.
	
	\item We embed failure recovery methods in clustering, application placement, and migration management techniques to recover from unpredicted failures.

\end{itemize}   
\par
The rest of paper is organized as follows. Relevant works of application placement and migration management techniques in edge and fog computing environments are discussed in section \ref{relatedw}. The system model and problem formulations are presented in section \ref{system}. Section \ref{placement} presents our proposed distributed clustering, application placement, and migration management technique. We evaluate the performance of our technique and compare it with the state-of-the-art techniques in section \ref{evaluation}. Finally, section~\ref{conclusion} concludes the paper and draws future works.
\section{Related Work}
\label{relatedw}
In this section, related works that address both application placement and mobility issues at the same time as their main challenges in the context of edge/fog computing are studied. These works are categorized into independent and dependent categories based on the dependency mode of their applications' granularity (e.g., modules). In the dependent category, constituent parts of IoT applications (i.e., modules) can be executed only when their predecessor modules complete their execution, while IoT applications that are modeled as a set of independent modules do not have this constraint.
\subsection{Edge Computing}			
In the independent category, Wang et al. \cite{wang2019dynamic} formulated service migration as a distance-based Markov Decision Process (MDP), which considers the distance between an IoT user and service provider as its main parameter. Then, they proposed a numerical technique to minimize the migration cost of users. Wang et al. \cite{wang2018user} and Yang et al \cite{yang2019efficient} considered deterministic mobility conditions, in which the potential paths between source and destination are priori known, and proposed placement techniques, performed on the IoT device, to minimize the delay. Since paths and available edge devices are priori known, as the IoT user moves, the current in-contact edge device can send the required information to the next edge device. Ouyang et al. \cite{ouyang2018follow} proposed an edge-centric application placement and mobility management technique that are executed on the network operator and one-hop edge devices respectively. They proposed a distributed approximation scheme based on the best response update technique to optimize the mobile edge service performance. Liu et al. \cite{liu2019mobility} proposed a mobility-aware offloading and migration technique to maximize the total revenue of IoT devices by reducing the probability of migration. Zhu et al. \cite{zhu2018fog} proposed a mobility-aware application placement in vehicular scenarios with constraints on service latency and quality loss. In this technique, some of the vehicles generate tasks while other vehicles provide computing services as remote servers. Zhang et al. \cite{zhang2019task} proposed a deep reinforcement technique to minimize the delay of IoT tasks. Yu et al. \cite{yu2018dmpo} proposed a technique to minimize the delay of tasks while satisfying the energy consumption of a single IoT user moving among edge servers. 
\par 
In the dependent category, Sun et al. \cite{sun2017emm} and Qi et al. \cite{qi2019knowledge} proposed a mobility-aware application placement technique in which placement decision engines run on IoT devices. The authors of \cite{sun2017emm} considered a single IoT device and proposed an IoT-centric energy-aware mobility management technique to minimize the application delay while authors of \cite{qi2019knowledge} proposed an edge-centric and knowledge-driven online learning method to adapt to the environment changes as vehicles move.			

\subsection{Fog Computing}
In the independent category, Wang et al.\cite{wang2016dynamic} proposed a solution to place a single service instance of each IoT user on a remote server when multiple IoT users exist in the system. They proposed both offline and online approximation algorithms, performed on the cloud, to find the optimal and near-optimal solutions respectively. Wang et al. \cite{wang2019mobility} and Wang et al. \cite{wang2019delay} proposed edge-centric application placement and mobility management technique when multiple IoT users with a single module exist in the system. The main goal of \cite{wang2019mobility} is Maximizing IoT users' gain through offloading and reducing the number of migrations, while the main goal of authors of \cite{wang2019delay} is minimizing the service delay.
\par
In the dependent category, Shekhar et al. \cite{shckhar2019urmila} and Bittencourt et al. \cite{bittencourt2017mobility} proposed mobility-aware application placement techniques for IoT application, consisting of multiple interdependent modules while considering prior mobility information. The authors in \cite{shckhar2019urmila} proposed a cloud-centric technique, called URMILA, in which the centralized controller makes the placement decision for all IoT applications to satisfy their latency requirements. Besides, whenever the decision is made, even in case the user leaves the range of its immediate server, there is no migration algorithm to migrate modules to new servers, which incurs a higher cost for the users. The authors in \cite{bittencourt2017mobility} proposed an edge-centric solution based on the edgeward-placement technique \cite{gupta2017ifogsim} for placement of IoT applications while considering their targeted destination. In this proposal, however, the potential of clustering is not considered. So, whenever the immediate server cannot serve the application modules, the modules are forwarded to the next hierarchical layer for possible placement and migration.

\subsection{A Qualitative Comparison}
\begin{table*}[!ht]
	\footnotesize
	\centering 
	\caption{A qualitative comparison of related works with ours}
	\label{tab:relatedwork}
	
	\resizebox{1\textwidth}{!}{%
		\renewcommand{\arraystretch}{1.5}
		\footnotesize
		\begin{tabular}{|c|c|c|c|c|c|c|c|c|c|c|c|c|c|} 
			\hline
			\multirow{3}{*}{Techniques}                                 & \multirow{3}{*}{Category}                                                  & \multicolumn{3}{c|}{\multirow{2}{*}{Application Properties}}                                                  & \multicolumn{3}{c|}{\multirow{2}{*}{Architectural Properties}}                                                                                                                                             & \multicolumn{6}{c|}{Placement and Mobility Management Engines}                                                                                                                                                                                                                                                            \\ 
			\cline{9-14}
			&                                                                            & \multicolumn{3}{c|}{}                                                                                         & \multicolumn{3}{c|}{}                                                                                                                                                                                      & \multirow{3}{*}{\begin{tabular}[c]{@{}c@{}} Placement Engine\\Position \end{tabular}} & \multirow{3}{*}{\begin{tabular}[c]{@{}c@{}} Mobility Management\\Engine Position \end{tabular}} & \multirow{3}{*}{\begin{tabular}[c]{@{}c@{}} Failure\\Recovery \end{tabular}} & \multicolumn{3}{c|}{Decision Parameters}                                            \\ 
			\cline{3-8}\cline{12-14}
			&                                                                            & Dependency                   & \begin{tabular}[c]{@{}c@{}}Module\\Number~\end{tabular} & Heterogeneity        & \begin{tabular}[c]{@{}c@{}}Number of\\IoT Devices \end{tabular} & \begin{tabular}[c]{@{}c@{}}Hierarchical\\Fog Architecture \end{tabular} & \begin{tabular}[c]{@{}c@{}}Clustering\\Technique~\end{tabular} &                                                                                       &                                                                                                 &                                 & Time                       & Energy                    & Weighted                   \\ 	\hline
			\cite{wang2019dynamic}                     & \multirow{10}{*}{\begin{tabular}[c]{@{}c@{}}Edge\\Computing \end{tabular}} & \multirow{8}{*}{Independent} & Single                                                  & Heterogeneous        & Single                                                          & $\times$                                               & $\times$                                      & Edge Centric                                                        & Edge Centric                                                                  & $\times$ & \checkmark  & $\times$ & $\times$  \\ 
			\cline{1-1}\cline{4-14}
			\cite{wang2018user}                        &                                                                            &                              & Multiple                                                & Heterogeneous        & Multiple                                                        & $\times$                                           & $\times$                                     & \begin{tabular}[c]{@{}c@{}}IoT Device\\Centric \end{tabular}        & Edge Centric                                                                  & $\times$ & \checkmark  & $\times$ &$\times$  \\ 
			\cline{1-1}\cline{4-14}
			\cite{ouyang2018follow}                    &                                                                            &                              & Single                                                  & Heterogeneous        & Multiple                                                        & $\times$                                               & $\times$                                     & Edge Centric                                                        & Edge Centric                                                                  &$\times$ & \checkmark & $\times$ & $\times$  \\ 
			\cline{1-1}\cline{4-14}
			\cite{liu2019mobility}                     &                                                                            &                              & Single                                                  & Heterogeneous        & Multiple                                                        & $\times$                                               & $\times$                                      & Edge Centric                                                        & Edge Centric                                                                  & $\times$ & \checkmark & \checkmark & \checkmark \\ 
			\cline{1-1}\cline{4-14}
			\cite{yang2019efficient}                   &                                                                            &                              & Multiple                                                & Heterogeneous        & Multiple                                                        & $\times$                                            & $\times$                                     & \begin{tabular}[c]{@{}c@{}}IoT Device\\Centric \end{tabular}        & Edge Centric                                                                  & $\times$ & \checkmark  & $\times$ & $\times$  \\ 
			\cline{1-1}\cline{4-14}
			\cite{zhu2018fog}                          &                                                                            &                              & Multiple                                                & Heterogeneous        & Single                                                          & $\times$                                               & $\times$                                     & Edge Centric                                                        & Edge Centric                                                                  &$\times$ & \checkmark  & $\times$ & $\times$  \\ 
			\cline{1-1}\cline{4-14}
			\cite{zhang2019task}                       &                                                                            &                              & Single                                                  & Homogeneous          & Single                                                          & $\times$                                              & $\times$                                   & Edge Centric                                                        & Edge Centric                                                                  & $\times$ & \checkmark & $\times$ & $\times$  \\ 
			\cline{1-1}\cline{4-14}
			\cite{yu2018dmpo}                          &                                                                            &                              & Multiple                                                & Heterogeneous        & Single                                                          & $\times$                                            & $\times$                                    & Edge Centric                                                        & Edge Centric                                                                  & $\times$ & \checkmark & \checkmark & $\times$ \\ 
			\cline{1-1}\cline{3-14}
			\cite{sun2017emm}                          &                                                                            & \multirow{2}{*}{Dependent}   & Multiple                                                & Heterogeneous        & Single                                                          & $\times$                                             & $\times$                                     & \begin{tabular}[c]{@{}c@{}}IoT Device\\Centric \end{tabular}        & \begin{tabular}[c]{@{}c@{}}IoT Device\\Centric \end{tabular}                  & $\times$ & \checkmark  & \checkmark &$\times$  \\ 
			\cline{1-1}\cline{4-14}
			\cite{qi2019knowledge}                     &                                                                            &                              & Multiple                                                & Heterogeneous        & Multiple                                                        &$\times$                                              & $\times$                                     & \begin{tabular}[c]{@{}c@{}}IoT Device\\Centric \end{tabular}        & Edge Centric                                                                  & $\times$ & \checkmark  & $\times$ & $\times$ \\ 
			\hline
			\cite{wang2016dynamic}                     & \multirow{6}{*}{\begin{tabular}[c]{@{}c@{}}Fog \\Computing \end{tabular}}  & \multirow{3}{*}{Independent} & Single                                                  & Heterogeneous        & Multiple                                                        & $\times$                                               & $\times$                                      & Cloud Centric                                                       & \begin{tabular}[c]{@{}c@{}}Cloud/Edge\\Centric \end{tabular}                  & $\times$ & \checkmark  & $\times$ & $\times$  \\ 
			\cline{1-1}\cline{4-14}
			\cite{wang2019mobility}                    &                                                                            &                              & Single                                                  & Heterogeneous        & Multiple                                                        & $\times$                                              & $\times$                                     & Edge Centric                                                        & Edge Centric                                                                  &$\times$ & \checkmark  & \checkmark &\checkmark  \\ 
			\cline{1-1}\cline{4-14}
			\cite{wang2019delay}                       &                                                                            &                              & Single                                                  & Heterogeneous        & Multiple                                                        & $\times$                                              & $\times$                                     & Edge Centric                                                        & Edge Centric                                                                  & $\times$ & \checkmark  & $\times$ & $\times$ \\ 
			\cline{1-1}\cline{3-14}
			\cite{shckhar2019urmila}                   &                                                                            & \multirow{3}{*}{Dependent}   & Multiple                                                & Homogeneous          & Single                                                          & $\times$                                               & $\times$                                      & Cloud Centric                                                       & Cloud Centric                                                                 & $\times$ & \checkmark  &$\times$ & $\times$  \\ 
			\cline{1-1}\cline{4-14}
			\cite{bittencourt2017mobility}             &                                                                            &                              & Multiple                                                & Heterogeneous        & Multiple                                                        & \checkmark                                              & $\times$                                     & Edge Centric                                                        & Edge Centric                                                                  & $\times$ & \checkmark  & $\times$ & $\times$  \\ 
			\cline{1-1}\cline{4-14}
			\begin{tabular}[c]{@{}c@{}}Proposed\\Solution \end{tabular} &                                                                            &                              & Multiple                                                & Heterogeneous        & Multiple                                                        & \checkmark                                               & \checkmark                                     & Edge Centric                                                        & Edge Centric                                                                  & \checkmark & \checkmark  & \checkmark & \checkmark  \\ 
			\hline
			\multicolumn{1}{c}{}                                        & \multicolumn{1}{c}{}                                                       & \multicolumn{1}{c}{}         & \multicolumn{1}{c}{}                                    & \multicolumn{1}{c}{} & \multicolumn{1}{c}{}                                            & \multicolumn{1}{c}{}                                                    & \multicolumn{1}{c}{}                                           & \multicolumn{1}{c}{}                                                & \multicolumn{1}{c}{}                                                          & \multicolumn{1}{c}{}      & \multicolumn{1}{c}{}       & \multicolumn{1}{c}{}      & \multicolumn{1}{c}{}       \\
			\multicolumn{1}{c}{}                                        & \multicolumn{1}{c}{}                                                       & \multicolumn{1}{c}{}         & \multicolumn{1}{c}{}                                    & \multicolumn{1}{c}{} & \multicolumn{1}{c}{}                                            & \multicolumn{1}{c}{}                                                    & \multicolumn{1}{c}{}                                           & \multicolumn{1}{c}{}                                                & \multicolumn{1}{c}{}                                                          & \multicolumn{1}{c}{}      & \multicolumn{1}{c}{}       & \multicolumn{1}{c}{}      & \multicolumn{1}{c}{}       \\
			\multicolumn{1}{c}{}                                        & \multicolumn{1}{c}{}                                                       & \multicolumn{1}{c}{}         & \multicolumn{1}{c}{}                                    & \multicolumn{1}{c}{} & \multicolumn{1}{c}{}                                            & \multicolumn{1}{c}{}                                                    & \multicolumn{1}{c}{}                                           & \multicolumn{1}{c}{}                                                & \multicolumn{1}{c}{}                                                          & \multicolumn{1}{c}{}      & \multicolumn{1}{c}{}       & \multicolumn{1}{c}{}      & \multicolumn{1}{c}{}       \\
			\multicolumn{1}{c}{}                                        & \multicolumn{1}{c}{}                                                       & \multicolumn{1}{c}{}         & \multicolumn{1}{c}{}                                    & \multicolumn{1}{c}{} & \multicolumn{1}{c}{}                                            & \multicolumn{1}{c}{}                                                    & \multicolumn{1}{c}{}                                           & \multicolumn{1}{c}{}                                                & \multicolumn{1}{c}{}                                                          & \multicolumn{1}{c}{}      & \multicolumn{1}{c}{}       & \multicolumn{1}{c}{}      & \multicolumn{1}{c}{}       \\
			\multicolumn{1}{c}{}                                        & \multicolumn{1}{c}{}                                                       & \multicolumn{1}{c}{}         & \multicolumn{1}{c}{}                                    & \multicolumn{1}{c}{} & \multicolumn{1}{c}{}                                            & \multicolumn{1}{c}{}                                                    & \multicolumn{1}{c}{}                                           & \multicolumn{1}{c}{}                                                & \multicolumn{1}{c}{}                                                          & \multicolumn{1}{c}{}      & \multicolumn{1}{c}{}       & \multicolumn{1}{c}{}      & \multicolumn{1}{c}{}      
		\end{tabular}		
	}	
	\vspace{-1.5cm}
\end{table*}

Key elements of related works are identified and presented in Table~\ref{tab:relatedwork} and compared with ours in terms of the main category, IoT application, architectural, and placement and mobility management engines' properties. The IoT application properties identify and compare dependency mode (either independent or dependent) of IoT applications, modules' number (either single or multiple modules per application), and heterogeneity (whether the specification of modules is same (i.e., homogeneous) or different (i.e., heterogeneous)). Architectural properties contain the number of IoT devices (either single or multiple), whether hierarchical fog architecture is considered or not, and clustering technique (whether a clustering technique is applied on edge/fog servers or not). Placement and mobility management engines contain positions of placement, mobility management engines, failure recovery capability, and the decision parameters used in each proposal.
\par
Our work proposes an edge-centric application placement and mobility management technique for an environment consisting of multiple IoT devices with heterogeneous applications (consisting of several dependent modules with heterogeneous requirements) and multiple remote servers (either CSs or FSs) deployed in a hierarchical architecture. Considering the potential of the clustering of FSs in the hierarchical fog computing environment, 
we propose a weighted cost model of response time and energy consumption for the application placement and migration techniques. The proposed weighted cost model considers the dependency among modules of IoT applications which plays an important role in application placement and migration management. Second, we put forward a distributed and dynamic clustering technique by which FSs of the same hierarchical level can form a cluster and collaboratively provide faster and more efficient service for IoT applications. This latter is because the communication overhead between FSs of the same hierarchical level is usually less than communication with higher-level FSs \cite{mahmud2018latency}. Although resources of each lower-level FS is less than each higher-level FS, aggregated resources of lower-level FSs, obtained through clustering, can be used to manage IoT applications modules in lower-level FSs with less response time and energy consumption. Third, we propose a distributed application placement and migration techniques for hierarchical fog computing environments to minimize the weighted cost of running real-time IoT applications. Finally, due to the highly dynamic nature of such systems, there is a high chance of failures in the system, for which we propose light-weight failure recovery methods in the clustering, application placement, and migration management techniques.

\section{System Model and Problem Formulation}
\label{system}
\begin{figure}[!t]
	\centering 
	\includegraphics[width=3.5in, height=7cm, trim=0.1in 0in 0in 0in]{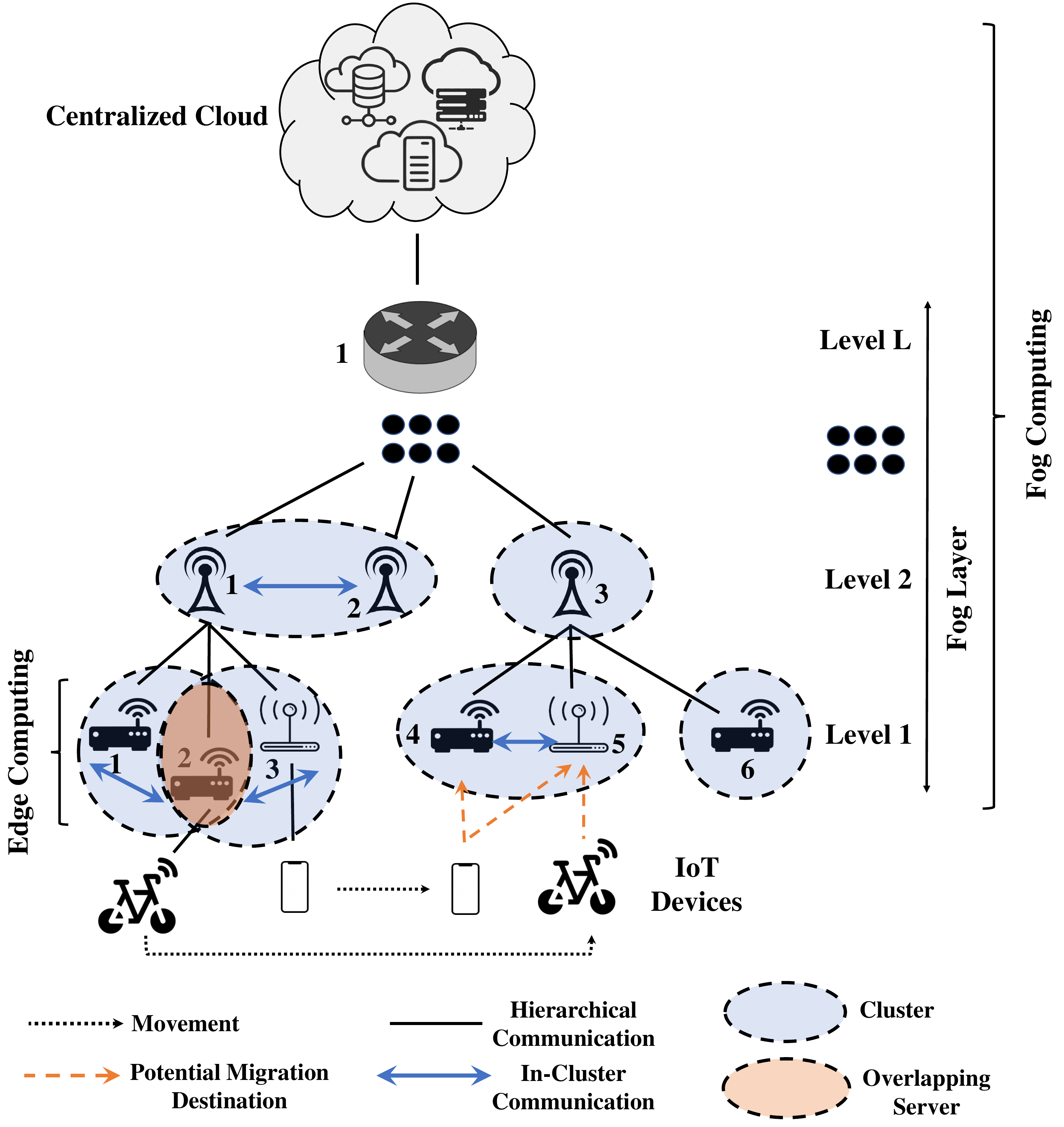}
	\caption{A view of our system model }
	\label{fig:systemmodel}
	\vspace{-.6cm}
\end{figure}

We consider a system consisting of $N$ mobile IoT users (so that each user has one IoT device), $F$ heterogeneous FSs distributed in the proximity of IoT users, and a centralized cloud. FSs follow a hierarchical topology, in which lower-level FSs can be accessed with lower latency while providing fewer resources in comparison to higher-level FSs that provide more resources but can be accessed with higher latency \cite{mahmud2018latency,pallewatta2019microservices}. Besides, we assume that each IoT device is connected to one FS in the lowest hierarchical level, so that this FS is responsible for the application placement and mobility management of that IoT device. The set of all available servers is represented as $\mathcal{S}$ with $|\mathcal{S}|=M$ and $M > F$. The 2-tuple $(h,i) \in \mathcal{S}$ ($0\leq h,1\leq i $) represents one server, in which $h$ represents the hierarchical level of the server and $i$ denotes the server's index at that hierarchical level. If we assume there are $L$ hierarchical fog layers, $(L+1,1)$ demonstrates the centralized cloud data-center placed at the top-most level. Moreover, the $(0,n)$ denotes the $n$th IoT device. Fig. \ref{fig:systemmodel} represents a view of our system model and how IoT devices move among different FSs. Also, it shows the in-cluster communications (in case clustering is applied) and communications between FSs at different hierarchical levels in this environment. 
\par  
%

\par
Each FS can form a cluster either by other nearby FSs at the same hierarchical level or by itself. Moreover, each FS in $l$th hierarchical level may belong to different clusters in that hierarchical layer. The cluster member (CM) list of each FS is defined as $List_{cl}(h,i)$, which is empty if the FS $(h,i)$ does not have any CMs. Besides, for each FS, we define a children list, $List_{ch}(h,i)$, containing server specification of immediate lower-level FSs, to which it has direct hierarchical communication links. The sole parent server of each FS is defined as $par(h,i)=(h^{\prime},i^{\prime})$ which refers to the immediate higher-level FS. We assume that in-cluster communications are faster than hierarchical communications \cite{mahmud2018latency}. Hence, clustering FSs, while incurs additional cost due to running clustering algorithm, can improve the quality of service for IoT users. Moreover, each FS has a list, called $\Omega(h,i)$, containing server specification of itself, its children, and all FSs belonging to the $\Omega$ of its children. To illustrate, considering Fig.~\ref{fig:systemmodel}, the $\Omega(2,1)=\{(2,1),(1,1),(1,2),(1,3)\}$  and $List_{ch}(2,1)=\{(1,1),(1,2),(1,3)\}$, and $\Omega(2,2)=\{(2,2)\} $ and $List_{ch}(2,2)=\{\}$. If we assume the maximum number of fog layers is three (i.e., $L=3$) in this example, then $\Omega(3,1)=\{(3,1),(2,1),(2,2),(2,3),(1,1),(1,2),(1,3), (1,4)\allowbreak,(1,5),(1,6)\}$, and the $List_{ch}(3,1)=\{(2,1),(2,2),(2,3)\}$. 
\par
We consider that FSs and CSs use container technology to run IoT applications' modules \cite{wang2019delay,sami2020vehicular}. So, we assume that FSs have access to images of all containers ($Cnts$) while such $Cnts$ may be active if they are running on the server or inactive (i.e., the container images are accessible, but the containers are not running) otherwise \cite{wang2019delay}. Moreover, for each container, according to the application module that it serves, an amount of ram size at the runtime is assigned to keep the state $Cnt_{v_{n,j}}^{ram}$ \cite{puliafito2019container}. Table~\ref{tab:notation} summarizes the parameters used in this paper and their respective definitions.
\begin{table*}[t]
	\centering
	\caption{Parameters and respective definitions}
	\label{tab:notation}
	\footnotesize
	\resizebox{1\textwidth}{!}{%
		\renewcommand{\arraystretch}{1.5}
		\begin{tabular}{|l|l|l|l|}
			\hline
			Parameter                                           & Definition                                                                                                                                                                   & Parameter                                         & Definition                                                                                                                                                                                                           \\ \hline
			CSs                                                 & Cloud Servers                                                                                                                                                                & FSs, FS                                           & Fog Servers, Fog Server                                                                                                                                                                                              \\ \hline
			$CNTs$, $CNT$                                       & Containers, Container                                                                                                                                                        & $N$                                               & Number of mobile IoT devices                                                                                                                                                                                         \\ \hline
			$F$                                                 & Number of heterogeneous fog servers (FSs)                                                                                                                                    & $\mathcal{S}$                                     & The set of all available servers                                                                                                                                                                                     \\ \hline
			$M$                                                 & Number of available servers                                                                                                                                                  & $(h,i)$                                           & \begin{tabular}[c]{@{}l@{}}The 2-tuple showing one server in which $h$ represents the hierarchical level of the\\  server and $i$ denotes the server's index at that hierarchical level\end{tabular}                 \\ \hline
			$List_{ch}(h,i)$                                    & The list containing server specification of children for the server $(h,i)$                                                                                                  & $par(h,i)$                                     & The sole parent of the server $(h,i)$ in the hierarchical system                                                                                                                                                     \\ \hline
			$\Omega(h,i)$                                       & \begin{tabular}[c]{@{}l@{}}The set containing server specification of server $(h,i)$, its children, and \\ all FSs belonging to the $\Omega$ of its children\end{tabular}    & CM                                                & Cluster Member                                                                                                                                                                                                      \\ \hline
			$List_{cl}(h,i)$                                    & The list containing server specification of cluster members for the server $(h,i)$                                                                                           & $G_n$                                             & Directed Acyclic Graph (DAG) of the $n$th IoT application                                                                                                                                                            \\ \hline
			$\mathcal{V}_n$                                     & The set of modules belonging to the $n$th IoT application                                                                                                                    & $\mathcal{E}_n$                                   & The set of data flows between modules belonging to the $n$th IoT application                                                                                                                                         \\ \hline
			$v_{n,i}$, $v_{n,j}$                                & The $i$th and/or $j$th module belonging to the $n$th IoT application                                                                                                         & $e_{n,i,j}$                                       & The data flow from module $v_{n,i}$ to module $v_{n,j}$ of the $n$th IoT device                                                                                                                                      \\ \hline
			$\mathcal{P}(v_{n,j})$                              & The set of predecessor modules of the module $v_{n,j}$                                                                                                                       & $TO_{n,i}=t$                                      & The topological order of $i$th module of the $n$th IoT application is equalt to $t$                                                                                                                                  \\ \hline
			$SchS_{n}$                                           & \begin{tabular}[c]{@{}l@{}}The schedule set of the $n$th IoT application consisting of subsets of \\ modules with the the same TO value $t$\end{tabular}                     & $SchS{n,t}$                                       & \begin{tabular}[c]{@{}l@{}}A subset of $SchS_{n}$ showing modules with the same TO value $t$ (i.e., modules \\ that can be executed in parallel)\end{tabular}                                                         \\ \hline
			$e_{n,i,j}^{ins}$                                   & \begin{tabular}[c]{@{}l@{}}The amount of instructions in terms of Million Instruction that the module\\  $v_{n,j}$ receives from $v_{n,i}$ for processing\end{tabular}       & $e_{n,i,j}^{dsize}$                               & \begin{tabular}[c]{@{}l@{}}The size of data that the module $v_{n,i}$ generates as an output to be sent to \\ module $v_{n,j}$\end{tabular}                                                                          \\ \hline
			$v_{n,i}^{mtd}$                                     & The maximum tolerable delay for the module $v_{n,i}$                                                                                                                         & $X_n$                                             & The placement configuration of the $n$th IoT application                                                                                                                                                             \\ \hline
			$x_{n,i}$                                           & \begin{tabular}[c]{@{}l@{}}The placement configuration for each module $v_{n,i}$ of the $n$th IoT\\  application in the $X_n$\end{tabular}                                   & $\Psi(X_n,t)$                                     & \begin{tabular}[c]{@{}l@{}}The weighted cost of modules in the $t$th schedule while considering the placement \\ configuration $X_n$.\end{tabular}                                                                   \\ \hline
			$|SchS_{n}|$                                         & The number of schedules for the $n$th IoT application                                                                                                                        & $T_{x_{n,j}}$                                     & The overall delay of each module (i.e., $v_{n,j}$) based on its assigned server                                                                                                                                      \\ \hline
			$Cnts_{(h,i)}$                                      & The number of instantiated $Cnts$ on the server $(h,i)$                                                                                                                      & $Cap_{(h,i)}$                                     & The maximum capacity of server $(h,i)$ to instantiate $Cnts$.                                                                                                                                                        \\ \hline
			$\Gamma(X_{n},t)$                                   & \begin{tabular}[c]{@{}l@{}}The weighted cost of modules in the $t$th schedule while considering the \\ placement configuration $X_n$\end{tabular}                            & $\Theta(X_n,t)$                                   & \begin{tabular}[c]{@{}l@{}}The energy consumption of modules in the $t$th schedule while considering the placement\\ configuration $X_n$\end{tabular}                                                                \\ \hline
			$T_{x_{n,j}}^{lat}$                                 & \begin{tabular}[c]{@{}l@{}}The inter-nodal latency between the servers on which module $v_{n,j}$ \\ and its predecessors $\mathcal{P}(v_{n,j})$ are placed\end{tabular}      & $T_{x_{n,j}}^{exe}$                               & \begin{tabular}[c]{@{}l@{}}The computing execution time of tasks, emitted from the $v_{n,i}$ to be \\ executed on the $v_{n,j}$\end{tabular}                                                                         \\ \hline
			$T_{x_{n,j}}^{tra}$                                 & \begin{tabular}[c]{@{}l@{}}The transmission time between between the module $v_{n,j}$ and its \\ predecessors $\mathcal{P}(v_{n,j})$\end{tabular}                            & $cpu(x_{n,j})$                                    & \begin{tabular}[c]{@{}l@{}}The computing power of the assigned server (in terms of MIPS) for the \\ module $v_{n,j}$\end{tabular}                                                                                    \\ \hline
			$\gamma^{tra}$                                      & The transmission time between source and destination servers                                                                                                                 & $B_{up}$, $B_{down}$, $B_{cluster}$               & \begin{tabular}[c]{@{}l@{}}The bandwidth of the one server to the parent server, to the child server, \\ and to its CMs, respectively\end{tabular}                                                                   \\ \hline
			$NST_{i}(H), NSE_{i}(H)$                            & They define the next intermediate server to reach the destination server                                                                                                     & $chRule$                                          & \begin{tabular}[c]{@{}l@{}}It identifies whether any children of the current server has a route to the \\ destination server or not\end{tabular}                                                                     \\ \hline
			$chRule$                                            & \begin{tabular}[c]{@{}l@{}}It identifies whether any CMs of the current server has a route to the \\ destination server or not\end{tabular}                                  & $\Upsilon((\Omega(h,i)),(h^{\prime},i^{\prime}))$ & \begin{tabular}[c]{@{}l@{}}It shows whether $\Omega(h,i)$ contains $(h^{\prime},i^{\prime})$ or not\\ (i.e., meaning that there is one hierarchical path from $(h,i)$ to the $(h^{\prime},i^{\prime})$)\end{tabular} \\ \hline
			$\gamma^{lat}$                                      & The inter-nodal latency between source and destination servers                                                                                                               & $lat(up)$, $lat(down)$, $lat(cluster)$            & \begin{tabular}[c]{@{}l@{}}The inter-nodal latency of one server to the parent server, to the child server,\\  and to its CMs, respectively\end{tabular}                                                             \\ \hline
			$\Psi^{mig}((X_n,X^{\prime}_{n},ts)$                    & \begin{tabular}[c]{@{}l@{}}The weighted migration cost of $n$th IoT application from the current \\ configuration $X_n$ to the new configuration $X^{\prime}_n$\end{tabular} & $\gamma^{mig}(x_{n,i},x_{n,i}^{\prime})$          & \begin{tabular}[c]{@{}l@{}}The migration cost of one module from current configuration $x_{n,i}$ to \\ the new configuration $x_{n,i}^{\prime}$\end{tabular}                                                         \\ \hline
			$\gamma_{mig}^{lat}((h,i),(h^{\prime},i^{\prime}))$ & The migration latency between current and new servers                                                                                                                        & $dsize^{mig}$                                     & \begin{tabular}[c]{@{}l@{}}The size of dump data and states that should be transferred between current \\ and new servers\end{tabular}                                                                               \\ \hline
			$e_{n,i,j}^{ins,r}$                                 & \begin{tabular}[c]{@{}l@{}}The amount of remaining instructions of task $e_{n,i,j}^{ins,r}$\\  to be executed on the new server after migration\end{tabular}                 & $E(x_{n,j})$                                      & The overall energy consumption of each module (i.e., $v_{n,j}$) based on its assigned server                                                                                                                         \\ \hline
			$E_{x_{n,j}}^{exe}$                                 & \begin{tabular}[c]{@{}l@{}}The computing energy consumption of tasks, emitted from the $v_{n,i}$to be\\ executed on the $v_{n,j}$\end{tabular}                               & $E_{x_{n,j}}^{lat}$                               & \begin{tabular}[c]{@{}l@{}}The energy consumption incurred due to inter-nodal latency between the servers on which \\ module $v_{n,j}$and its predecessors $\mathcal{P}(v_{n,j})$ are placed\end{tabular}            \\ \hline
			$E_{x_{n,j}}^{tra}$                                 & \begin{tabular}[c]{@{}l@{}}The transmission energy consumption between between the module $v_{n,j}$\\ and its predecessors $\mathcal{P}(v_{n,j})$\end{tabular}               & $P_{cpu}$, $P_{i}$, $P_{t}$                       & \begin{tabular}[c]{@{}l@{}}The CPU power of the IoT device, the idle power of IoT device, and transmission power of\\ the IoT device\end{tabular}                                                                    \\ \hline
			$\vartheta^{tra}$                                   & The transmission energy consumption between source and destination servers                                                                                                   & $\vartheta^{lat}$                                 & The energy consumption incurred due to inter-nodal latency between servers                                                                                                                                           \\ \hline
			$\Gamma^{mig}((X_n,X^{\prime}_{n}),t)$                  & \begin{tabular}[c]{@{}l@{}}The migration time of $n$th IoT application from the current configuration\\ $X_n$ to the new configuration $X^{\prime}_n$ considering schedule $t$\end{tabular}           & $\Theta^{mig}((X_n,X^{\prime}_{n}),t)$                & \begin{tabular}[c]{@{}l@{}}The migration energy consumption of $n$th IoT application from the current configuration\\ $X_n$ to the new configuration $X^{\prime}_n$ considering schedule $t$\end{tabular}                                     \\ \hline
		\end{tabular}
	}
\vspace{-3mm}
\end{table*}


\subsection{Application Model}
We consider real-time IoT applications working based on the Sense-Process-Actuate model, in which sensors transmit tasks periodically according to their sample rate \cite{mahmud2018latency,gupta2017ifogsim}. The emitted sensors' tasks should be forwarded to different modules of the IoT applications for processing based on dependency model among constituent modules. When each module receives tasks from predecessor modules as input, it processes tasks and produces respective tasks as its output to be forwarded to next modules \cite{mahmud2018latency}. Finally, results will be forwarded to the actuator as the last module. In this work, we assume that both sensor and actuator modules of IoT applications reside in IoT devices \cite{bittencourt2017mobility}.
\par
\begin{figure*}[!t]
	\centering
	\begin{subfigure}{0.49\textwidth}
		\includegraphics[height=4cm, trim=0.1in 0in 0in 0in]{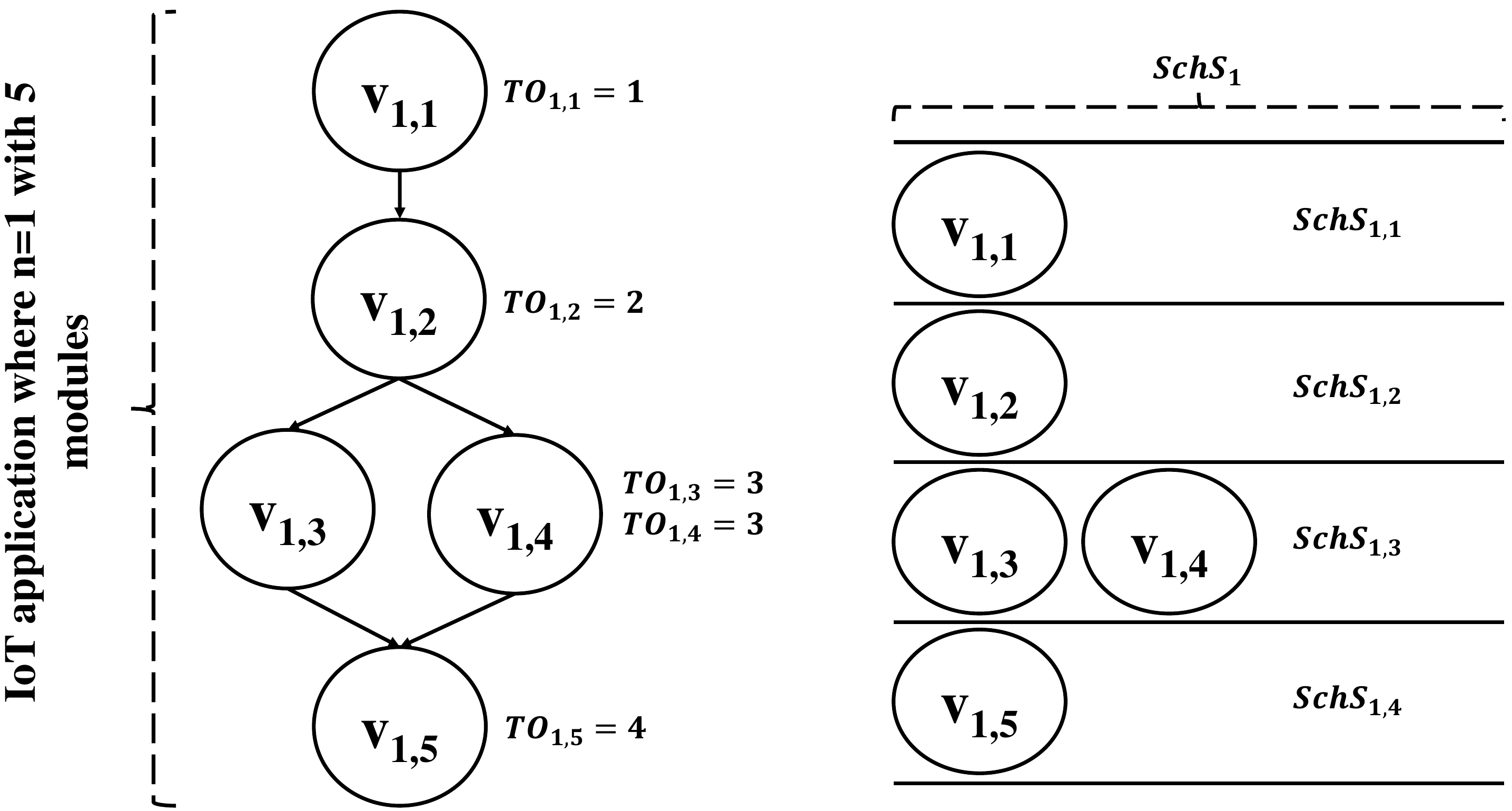}
		\caption{An IoT application and its corresponding schedules}
		\label{fig:TopologicalOrder}
	\end{subfigure}
	\begin{subfigure}{0.49\textwidth}
		\vspace{2mm}
		\includegraphics[height=4cm, trim=0.1in 0in 0in 0in]{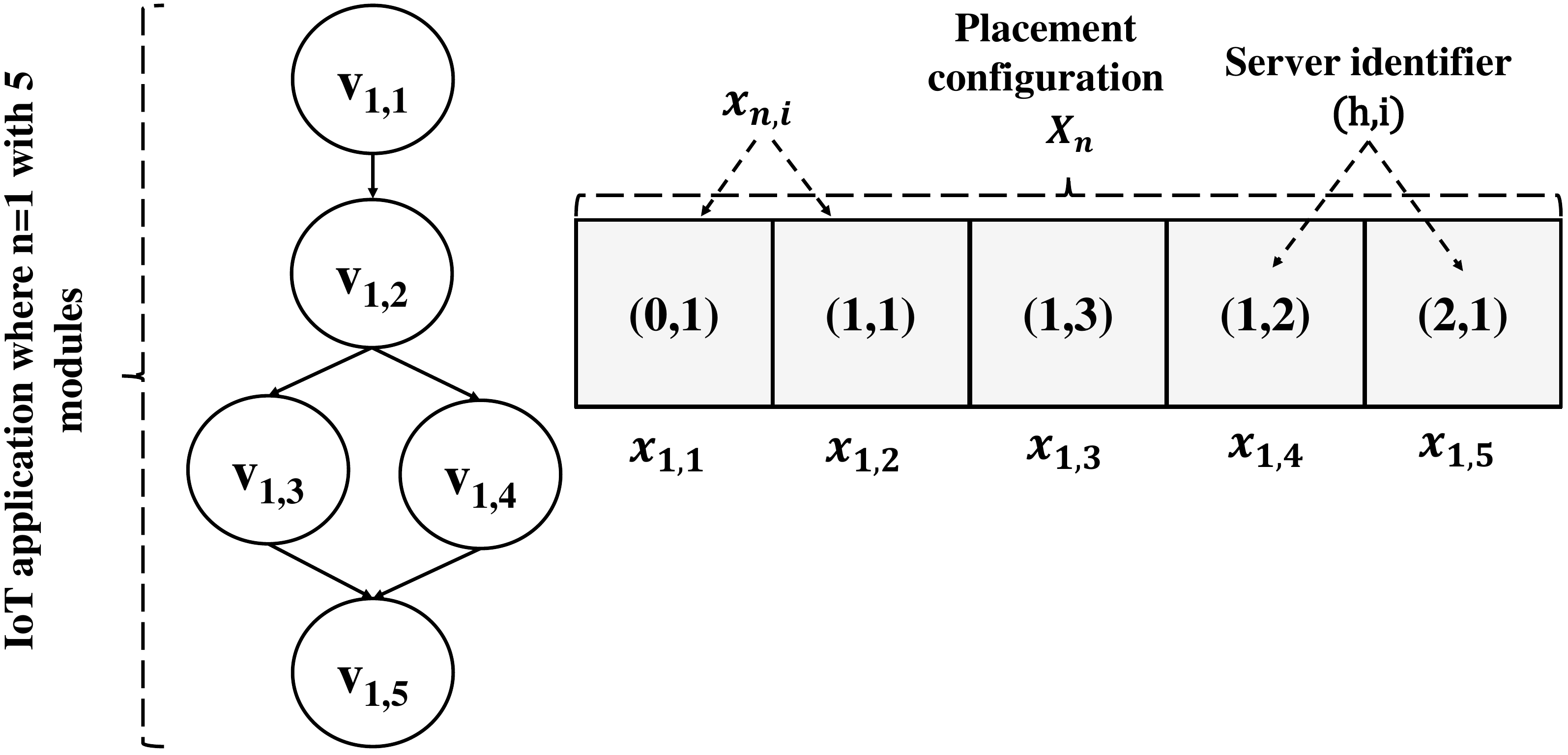}
		\caption{A candidate server configuration for the IoT application}
		\label{fig:SolutionConfig}
	\end{subfigure}
	\caption{An example of IoT application, its schedules and a candidate server configuration}
	\label{fig:IoTApplicationExample}
	\vspace{-0.5cm}
\end{figure*}

   
Real-time IoT application belonging to the $n$th IoT device is represented as a Directed Acyclic Graph (DAG) of its modules $G_n=(\mathcal{V}_n,\mathcal{E}_n),\forall n\in\{1,2,\cdots,N\}$, where $\mathcal{V}_n=\{v_{n,i}|1\leq i \leq |\mathcal{V}_n|\}$ denotes the set of modules belonging to the $n$th IoT device, and $\mathcal{E}_n=\{e_{n,i,j}|v_{n,i}, v_{n,j} \in \mathcal{V}_n,\thinspace v_{n,i} \in \mathcal{P}(v_{n,j}), \thinspace i \neq j\}$ shows the set of data flows between modules. Since IoT applications are modeled as DAGs, each module $v_{n,j}$ cannot be executed unless all its predecessor modules, denoted as $\mathcal{P}(v_{n,j})$, finish their execution. To illustrate, $e_{1,1,2}$ represents that execution of module $v_{1,2}$ depends on the execution of the module $v_{1,1}$. Moreover, we define a Topological Order value $t$ for each module $i$ of the $n$th IoT application as $TO_{n,i}=t$. We define a schedule set for the $n$th IoT application, called $SchS_{n}$, consisting of modules with the same TO value $t$ as its subsets. The $SchS_{n,t}$ specify modules with the same TO value $t$ (i.e., modules that can be executed in parallel). In addition, the set of successor modules of module $v_{n,j}$ is defined as $Succ(v_{n,j})$. Fig~\ref{fig:TopologicalOrder} shows an IoT application, the TO value for each module, and the schedule set $SchS_{n}$ based on the TO values of its modules.
Besides, We define the output of each module $v_{n,i}$ is a task consisting of two values to be forwarded to next modules based on data flows of the IoT application. The first value is the amount of instructions in terms of Million Instruction (MI) that the module $v_{n,j}$ receives from $v_{n,i}$ for processing, shown as $e_{n,i,j}^{ins}$, and the second value is the size of data $e_{n,i,j}^{dsize}$ the module $v_{n,i}$ generates as its output to be forwarded to module $v_{n,j}$ \cite{gupta2017ifogsim}. 

\subsection{Problem Formulation}
The placement configuration of the application belonging to the $n$th IoT application is shown as $X_n$. Also, $x_{n,i}=(h,i)$ denotes the placement configuration for each module $v_{n,i}$ of the $n$th IoT application in the $X_n$ based on the specification of the server. To illustrate, $x_{n,i}= (1,3)$ shows that the $i$th module of $n$th IoT device is assigned to a server in the first hierarchical level where the server index is $3$. Moreover, if the $i$th module of the $n$th IoT device is assigned to run locally on itself, $x_{n,i}=(0,n)$. Fig~\ref{fig:SolutionConfig} presents a sample DAG of an IoT application and a candidate placement configuration. 

\subsubsection{Placement weighted cost model}
The goal of application placement is to find a suitable configuration for modules of each real-time IoT application to minimize the weighted cost $\Psi(X_n,t)$ of running applications in terms of the response time of tasks and energy consumption of IoT devices:
\begin{equation}
	\min\limits_{w_{1},w_{2} \in [0,1]}\sum\limits_{t=1}^{|SchS_{n}|} \Psi(X_n,t), \quad \forall n \in \{1,2,\cdots,N\}
	\label{weightOmega}
\end{equation}

where
\begin{equation}
	\Psi(X_n,t) = w_{1} \times \Gamma(X_n,t) + w_{2} \times \Theta(X_n,t)
	\label{controlOmega}
\end{equation}

\begin{eqnarray}
	\hspace{-1cm}
	s.t.&&C1:\;Size(x_{n,j})=1,\; \forall x_{n,j} \in X_n \; , \\
	&&\; \;\;\;\;\;\;\;\; n \in \{1,2,\cdots,N\}, 1\leq i \leq |\mathcal{V}_n|  \nonumber\\
	&&C2:\; Cnts(h,i) \leq Cap(h,i),\; \forall \; (h,i) \in \mathcal{S}\\
	&&C3:\;\Psi(x_{n,i},t) \leq \Psi(x_{n,j},t),\; \forall v_{n,i}\in\mathcal{P}(v_{n,j})   
\end{eqnarray}
\noindent
where $|SchS_{n}|$ represents the number of schedules, and $\Gamma(X_{n},t)$ and $\Theta(X_{n},t)$ show the response time model and energy consumption model, respectively, of modules in the $t$th schedule while considering the placement configuration $X_n$. Moreover, $w_1$ and $w_2$ are control parameters to tune the weighted cost model according to user requirements. We assume the number of available servers $M$ is more than or equal to the maximum number of modules in the $t$th schedule for parallel execution (i.e., $|SchS_{n,t}|\leq M$). We suppose that each module of an IoT application can be exactly assigned to one $Cnt$ of one remote server. $C1$ indicates that each module $i$ of the $n$th IoT application can only be assigned to one server at a time, and hence the size of $x_{n,j}$ is equal to 1 \cite{goudarzi2020application,xu2019computation}. $C2$ denotes that the number of instantiated $Cnts$ on the server $(h,i)$ is less or equal to the maximum capacity of that server $Cap(h,i)$. Besides, $C3$ guarantees that the predecessor modules of $v_{n,j}$ (i.e., $\mathcal{P}(v_{n,j})$) are executed before the execution of module $v_{n,j}$ \cite{xu2019computation}.
\paragraph{\textbf{Response time model}}
The goal of this model is to find the best possible configuration of servers for each IoT application so that the overall response time for each IoT application becomes minimized. In order to only consider response time model as the main objective, the control parameters of weighted cost model (Eq.~\ref{controlOmega}) can be set to $w_1=1$ and $w_2=0$.

\begingroup
\footnotesize
\begin{equation}
	\label{eq.ParallelSequential}
	\Gamma(X_n,t)=\left\{ \begin{tabular}{ll}
		$T(x_{n,j})$, \;\;\;\;\;\;\;\;\;\;\;\;\; \text{if }$|SchS_{n,t}|=1$  &(a)\vspace{0.4cm}\\
		$max(T(x_{n,j}))$, \;\;\;\; otherwise\\
		& (b)\\
		$\forall x_{n,j}\in X_{n}|v_{n,j}\in SchS_{n,t}$
		
	\end{tabular}\right.
\end{equation}
\endgroup

The Eq.~\ref{eq.ParallelSequential}.a represents the condition in which the number of modules in the $t$th schedule is one (i.e, $|SchS_{n,t}|=1$), and hence, the time of that schedule is equal to the time of that module based on its assigned server $T(x_{n,j})$. Besides, the Eq.~\ref{eq.ParallelSequential}.b refers to the condition in which the number of modules in the $t$th schedule is more than one (i.e., several modules can be executed in parallel). In this latter case, the time of the $t$th schedule is equal to the maximum time of all modules that can be executed in parallel.

\par
The overall delay of each module (i.e., $v_{n,j}$) based on its candidate configuration (i.e., $x_{n,j}$) is defined as the sum of inter-nodal latency between servers ($T_{x_{n,j}}^{lat}$), the computing time per module ($T_{x_{n,j}}^{exe}$), and the data transmission time between $v_{n,j}$ and all of its predecessor modules ($T_{x_{n,j}}^{tra}$). It is formulated as:
\begin{equation}
	\label{equation:totalTime}
	T(x_{n,j}) = T_{x_{n,j}}^{exe} + T_{x_{n,j}}^{lat} + T_{x_{n,j}}^{tra}
\end{equation}
\noindent
The computing execution time of module $v_{n,j}$ depends on tasks emitted from its predecessors (i.e., $\mathcal{P}(v_{n,j})$) for processing by $v_{n,j}$. The computing time of $v_{n,j}$ is estimated as:

\begin{eqnarray} 
	\label{equation:totalTimeExeModule}
	T_{x_{n,j}}^{exe}=\sum\frac{e_{n,i,j}^{ins}}{cpu(x_{n,j})},\\
	\forall e_{n,i,j}\in \mathcal{E}_n|v_{n,i}\in \mathcal{P}(v_{n,j}), \nonumber
\end{eqnarray}

\noindent
where $cpu(x_{n,j})$ demonstrates the computing power of the assigned server (in terms of Million Instruction per Second (MIPS)) for the module $v_{n,j}$. Moreover,  the $e_{n,i,j}^{ins}$ shows the amount of instructions in terms of MI that the module $v_{n,j}$ receives from $v_{n,i}$ for the processing.
\par
The transmission time between module $v_{n,j}$ and its predecessors $\mathcal{P}(v_{n,j})$ of the application belonging to the $n$th IoT device is calculated as:
\begin{eqnarray}
	T_{x_{n,j}}^{tra}=\max (\gamma^{tra}(e_{n,i,j}^{dsize},(h,i),(h^{\prime},i^{\prime}))),\\
	\forall e_{n,i,j}\in \mathcal{E}_n|v_{n,i}\in \mathcal{P}(v_{n,j}), \nonumber\\
	x_{n,i}=(h,i),x_{n,j}=(h^{\prime},i^{\prime})  \nonumber
\end{eqnarray}
\noindent
Due to the hierarchical nature of fog computing, the transmission time of one task ($\gamma^{tra}$) between each pair of dependent modules $v_{n,i}$ and $v_{n,j}$ is recursively obtained based on visited servers between source and destination. The $(h,i)$ and $(h^{\prime},i^{\prime})$ show server specifications of source and destination servers on which modules $v_{n,i}$ and $v_{n,j}$ are assigned, respectively. By visiting each intermediate server between source and destination servers, the value of source server $(h,i)$ is updated while the value of destination server remains unchanged. To reduce the length of equations, we consider $(e_{n,i,j}^{dsize},(h,i),(h^{\prime},i^{\prime}))=H$.

\begingroup
\footnotesize
\begin{equation}
	\hspace{-14mm}
	\label{eq.NSParentTransmission}
	\gamma^{tra}(H)= \left\{ \begin{tabular}{ll} 
		$\frac{e_{n,i,j}^{dsize}}{B_{up}}+\gamma^{tra}(H^{\prime}),$&$NST_{i}(H)=NST_{1}|NST_{4}|NST_{6}$\\
		$\frac{e_{n,i,j}^{dsize}}{B_{down}}+\gamma^{tra}(H^{\prime}),$ & $NST_{i}(H)=NST_{2}$\vspace{0.1cm}\\
		$\frac{e_{n,i,j}^{dsize}}{B_{cluster}}+\gamma^{tra}(H^{\prime}),$ & $NST_{i}(H)=NST_{3}|NST_{5}$\vspace{0.1cm}\\
		$0$,& $NST_{i}(H)=NST_{7}$
	\end{tabular}\right.
\end{equation}
\endgroup

\noindent
where $B_{up}$, $B_{down}$, and $B_{cluster}$ refer to the bandwidth of current server to parent server, to child server, and to cluster server, respectively. Besides, $H^{\prime}$ is defined as what follows:
\begin{equation}
	\label{eq.Hprime}
	H^{\prime}=(e_{n,i,j}^{dsize},(h^{''},i^{''}),(h^{\prime},i^{\prime}))
\end{equation}
\begin{equation}
	\label{eq.Huuuura}
	(h^{''},i^{''})=NST_{i}(H)
\end{equation}
The Eq.~\ref{eq.Hprime} shows the data size and destination server of $H^{\prime}$ is exactly the same as $H$, and the only difference is the specification of the source server $(h^{''},i^{''})$ which is obtained from the output of $NST_{i}(H)$ (i.e., $(h^{''},i^{''})=NST_{i}(H)$. The $NST_{i}(H)$ defines the next intermediate server to reach the destination server for each edge $e_{n,i,j}$.

\begingroup
\footnotesize
\begin{eqnarray}
	NST_{i}(H)= \left \{ \begin{array}{llc}
		Par(h,i), &\text{if } h < h^{\prime} \vspace{3mm}&i=1\\
		
		chRule,&\text{if } h> h^{\prime} &i=2\\ &\&\hspace{1mm} chRule\neq \varnothing\vspace{3mm}\\
		
		&\text{if } h\varoplus h^{\prime}=0    \\clRule, &\&\hspace{1mm}  i \varoplus i^{\prime}\neq 0   &i=3     \\  &\&\hspace{1mm}  clRule\neq \varnothing\vspace{3mm}\\
		
		&\text{if } h\varoplus h^{\prime}=0    \\Par(h,i), &\&\hspace{1mm}  i \varoplus i^{\prime}\neq 0   &i=4     \\  &\&\hspace{1mm}  clRule= \varnothing\vspace{3mm}\\ 	 
		
		clRule,&\text{if } h> h^{\prime},&i=5 \\ &\&\hspace{1mm} clRule\neq \varnothing\vspace{4mm}\\

		&\text{if } h>h^{\prime}    \\Par(h,i), &\&\hspace{1mm}    chRule=\varnothing   &i=6   \\  &\&\hspace{1mm}  clRule=\varnothing\vspace{3mm}\\
		
		(0,0),&\text{if } h\varoplus h^{\prime}=0&i=7 \\ &\&\hspace{1mm} i\varoplus i^{\prime}=0
		
	\end{array}\right.
	\label{NScontrol}
\end{eqnarray}
\endgroup

\begin{eqnarray}
	\label{chrule}
	chRule=\textrm{if } \exists (h^{''},i^{''}) \in List_{ch}(h,i) |\nonumber\\
	\Upsilon((\Omega(h^{''},i^{''}),(h^{\prime},i^{\prime}))=1,
	\textrm{return } (h^{''},i^{''}), \\
	\textrm {else return $\varnothing$}\nonumber
\end{eqnarray}

\begin{eqnarray}
	\label{clRule}
	clRule=\textrm{if } \exists (h^{''},i^{''}) \in List_{cl}(h,i) |\nonumber\\ \Upsilon((\Omega(h^{''},i^{''}),(h^{\prime},i^{\prime}))=1, \textrm{return } (h^{''},i^{''}), \\
	\textrm {else return $\varnothing$}\nonumber 
\end{eqnarray}

\begin{figure*}[!ht]
	\begin{subfigure}{.325\textwidth}
		\centering
		\includegraphics[width=\linewidth,height=7cm]{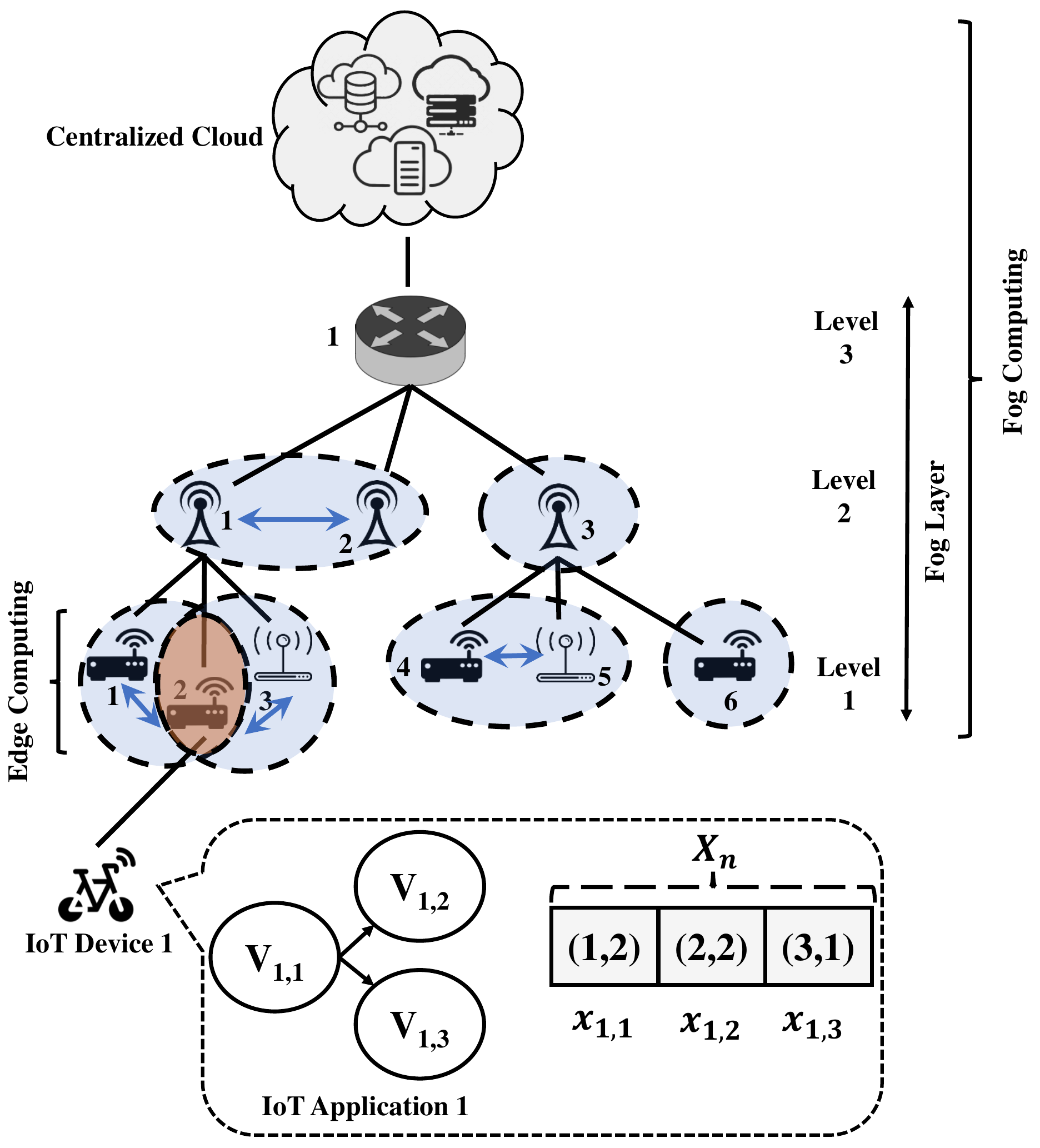}
		\captionsetup{justification=centering}
		\subcaption{An IoT application and its candidate configuration}
		\label{fig:exampleTransmission:a}
	\end{subfigure}%
	\hspace{0.1cm}
	\begin{subfigure}{.325\textwidth}
		\centering
		\includegraphics[width=\linewidth,height=7cm]{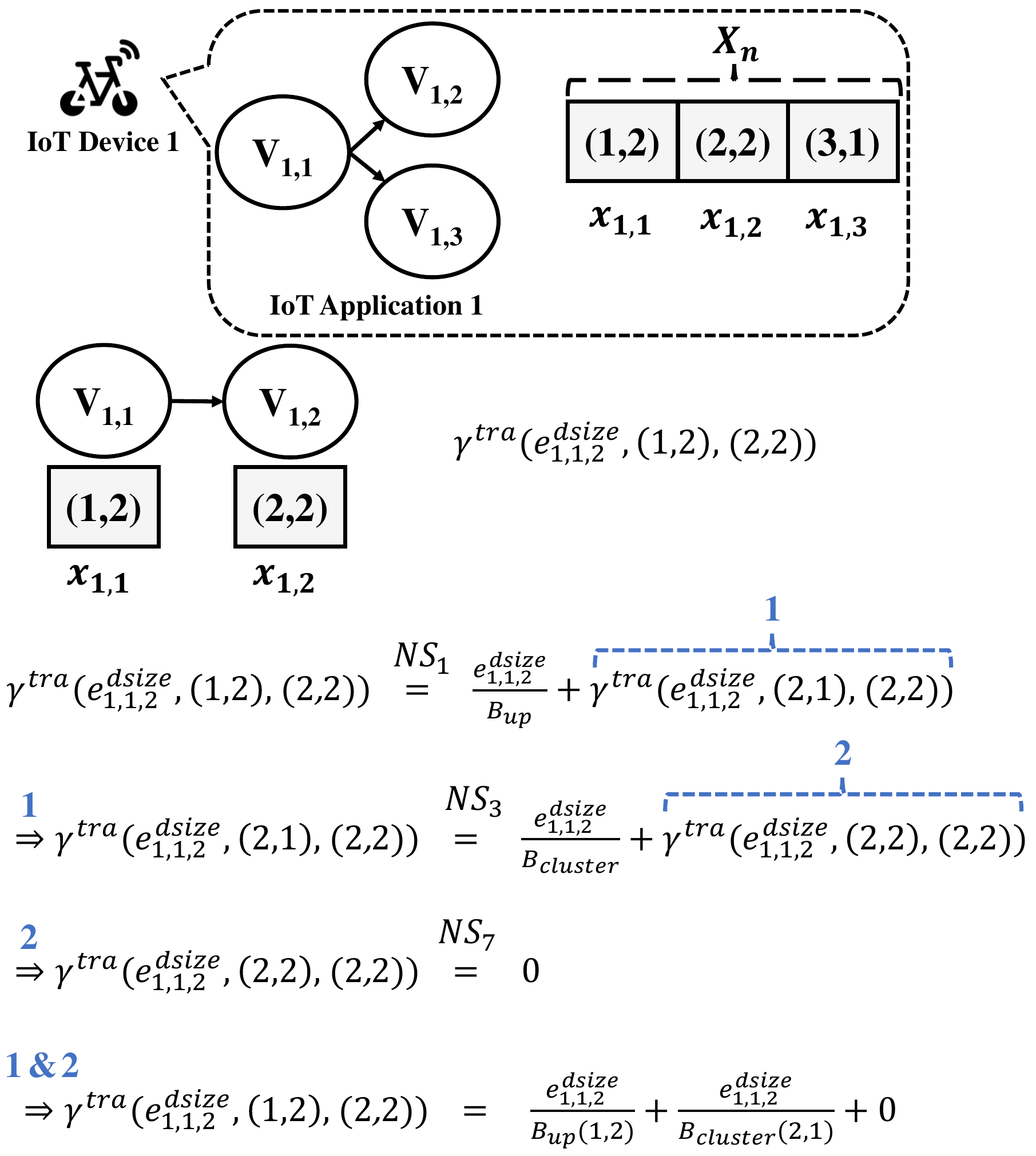}
		\captionsetup{justification=centering}
		\subcaption{Transmission time for the data flow $e_{1,1,2}$}
		\label{fig:exampleTransmission:b}
	\end{subfigure}
	\hspace{0.25mm}
	\begin{subfigure}{.325\textwidth}
		\centering
		\includegraphics[width=\linewidth,height=7cm]{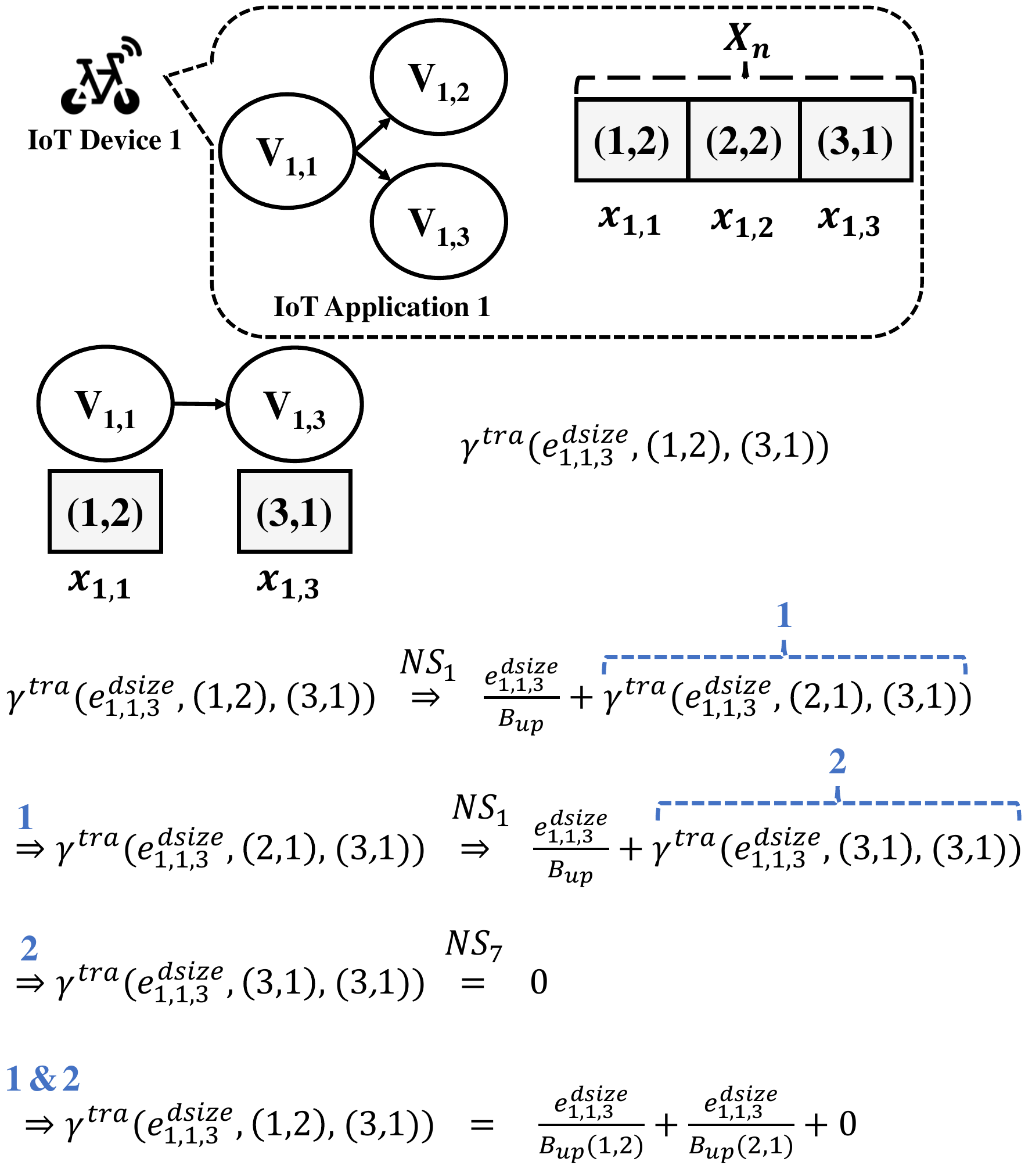}
		\captionsetup{justification=centering}
		\subcaption{Transmission time for the data flow $e_{1,1,3}$}
		\label{fig:exampleTransmission:c}
	\end{subfigure}
	
		\caption{A example of calculating transmission time based on a candidate configuration}
		\label{fig:exampleTransmission}
\end{figure*}

\noindent
The $\Upsilon((\Omega(h^{''},i^{''})),(h^{\prime},i^{\prime}))$ is equal to $1$ if $\Omega(h^{''},i^{''})$ contains $(h^{\prime},i^{\prime})$ (i.e., meaning that there is one hierarchical path from $(h^{''},i^{''})$ to the $(h^{\prime},i^{\prime})$) and is equal to $0$ if $(h^{\prime},i^{\prime})$ does not exist. Moreover, the $\varoplus$ is XOR binary operation. The chRule (Eq.~\ref{chrule}) says that if the server $(h,i)$ has a children $(h^{''},i^{''})$ in its $List_{ch}$ which has a hierarchical path to the destination server $(h^{'},i^{'})$, the specification of this server $(h^{''},i^{''})$ should be returned. The clRule (Eq.~\ref{clRule}) presents that if the server $(h,i)$ has a CM $(h^{''},i^{''})$ in its $List_{cl}(h,i)$ which has a hierarchical path to the destination server $(h^{'},i^{'})$, the specification of this server $(h^{''},i^{''})$ should be returned. Based on the aforementioned rules, $NST(H)$ finds the next server to which the data should be sent and calculates the transmission cost. The $NST_{1}$ of Eq.~\ref{NScontrol} states that if the hierarchical level of the current server is less than destination server, the $Par(h,i)$ should be checked in the next step. The $NST_{2}$ represents the case that the hierarchical level of the current server is higher than the destination server and the current server has a child through which the destination server can be reached. The $NST_{3}$ states the condition that the current and destination servers are in the same hierarchical level, and one of the CMs has a route to the destination server. The $NST_{4}$ indicates that if the current and the destination servers are in the same level, and there is no route to destination using CMs, the parent should be checked in the next step. The $NST_{5}$ states that if the level of the current server is higher than the destination server, and a CM has a path to the destination server, the cluster server should be selected in the next step. The $NST_{6}$ states that if the level of current server is higher than the destination server, and there exists no route from children nor from CMs, the parent server should be traversed. Finally, the $NST_{7}$ is the ending condition for this recursive process and states that if the current and destination server is same, the cost is zero. Fig~\ref{fig:exampleTransmission} represents an example of obtaining transmission time between source and destination servers.
%

\noindent
Inter-nodal latency $T_{x_{n,j}}^{lat}$ between servers on which module $v_{n,j}$ and its predecessors $\mathcal{P}(v_{n,j})$ are placed is calculated as:

\begin{eqnarray} 
	\Gamma_{X_{n,j}}^{lat}=\max  (\gamma^{lat}((h,i),(h^{\prime},i^{\prime}))), \\
	\forall e_{n,i,j}\in \mathcal{E}_n|v_{n,i}\in \mathcal{P}(v_{n,j}), \nonumber\\
	x_{n,i}=(h,i),x_{n,j}=(h^{\prime},i^{\prime})  \nonumber
\end{eqnarray}

\noindent
where $\gamma^{lat}$ shows the inter-nodal latency between source and destination servers (i.e., $(h,i)$ and $(h^{\prime},i^{\prime})$ respectively) on which $v_{n,i}$ and $v_{n,j}$ are placed. It is calculated similar to the transmission time. To reduce the equation size, we consider $((h,i),(h^{\prime},i^{\prime}))=A$.
\begingroup
\footnotesize
\begin{equation}
	\hspace{-14mm}
	\label{eq.NSParentLatency}
	\gamma^{lat}(A)= \left\{ \begin{tabular}{ll} 
		$lat_{up}+\gamma^{lat}(A^{\prime})$, & $NST_{i}(A)=NST_{1}|NST_{4}|NST_{6}$ \vspace{0.1cm}\\
		$lat_{down}+\gamma^{lat}(A^{\prime})$, & $NST_{i}(A)=NST_{2}$\vspace{0.1cm}\\
		$lat_{cluster}+\gamma^{lat}(A^{\prime})$, & $NST_{i}(A)=NST_{3}|NST_{5}$\vspace{0.1cm}\\
		$0$,& $NST_{i}(A)=NST_{7}$
	\end{tabular}\right.
\end{equation}
\endgroup

\noindent
where $lat_{up}$, $lat_{down}$, and $lat_{cluster}$ correspond to up-link, down-link, and cluster-link inter-nodal latency respectively, and depends on the hierarchical level of servers. Besides, $A^{\prime}$ is defined as what follows:
\begin{equation}
	\label{eq.Aprime}
	A^{\prime}=((h^{''},i^{''}),(h^{\prime},i^{\prime}))
\end{equation}
The Eq.~\ref{eq.Aprime} shows the destination server (i.e., $(h^{\prime},i^{\prime})$) of $A^{\prime}$ is exactly the same as $A$, and the only difference is the specification of the source server $(h^{''},i^{''})$ which is obtained from the output of $NST(A)$. The $NST(A)$ performs exactly the same as $NST(H)$ (i.e., Eq.~\ref{NScontrol}) to find the next intermediate server, and all equation from Eq.~\ref{NScontrol} to Eq.~\ref{clRule} are valid here.
\paragraph{\textbf{Energy consumption model}}
The goal of this model is to find a suitable placement configuration of application modules to minimize the energy consumption of the $n$th IoT device. To only consider energy consumption model as the main objective, the control parameters of weighted cost model (Eq.~\ref{controlOmega}) can be set to $w_1=0$ and $w_2=1$.

\begingroup
\footnotesize
\begin{equation}
	\label{eq.EnergyParallelSequential}
	\Theta(X_n,t)=\left\{ \begin{tabular}{ll}
		$E(x_{n,j})$, \;\;\;\;\;\;\;\;\;\;\;\;\; \text{if }$|SchS_{n,t}|=1$  &(a)\\
		\\
		$max(E(x_{n,j}))$, \;\;\;\; otherwise\\
		& (b)\\
		$\forall x_{n,j}\in X_{n}|v_{n,j}\in SchS_{n,t}$
		
	\end{tabular}\right.
\end{equation}
\endgroup

\noindent
where $|SchS_{n}|$ shows the number of schedules, and $\Theta(X_{n},t)$ represents the energy consumption of modules in the $t$th schedule while considering the placement configuration $X_n$.
\par
The overall energy consumption of each module (i.e., $v_{n,j}$) based on its candidate configuration (i.e., $x_{n,j}$) is defined as the sum of energy consumed for inter-nodal latency between servers ($E_{x_{n,j}}^{lat}$), the computing of each module ($E_{x_{n,j}}^{exe}$), and the data transmission between $v_{n,j}$ and all of its predecessor modules ($E_{x_{n,j}}^{tra}$). It is formulated as:
\begin{equation}
	\label{equation:totalEnergy}
	E(x_{n,j}) = E_{x_{n,j}}^{exe} + E_{x_{n,j}}^{lat} + E_{x_{n,j}}^{tra}
\end{equation}

The computing energy consumption for module $v_{n,j}$ depends on its assigned server and can be derived from:

\begin{equation}
	\hspace{-2mm}
	\label{equation:totalEnergyExeModule}	
	E_{x_{n,j}}^{exe}= \left\{ \begin{tabular}{ccc} $T_{x_{n,j}}^{exe}\times P_{cpu}$, &\text{if }$x_{n,j}=(h,i) \;\&\; h=0$ \vspace{0.1cm}\\
		$T_{x_{n,j}}^{idle}\times P_{i}$, &\text{if } $x_{n,j}=(h,i) \;\&\; h\neq0$
	\end{tabular}\right.
\end{equation}

Because only the energy consumption of IoT devices is considered in this work, whenever application modules run on remote servers, the energy consumption of IoT device is equal to the idle time $T_{x_{n,j}}^{idle}$ multiplied to the power consumption of IoT device in its idle mode $P_{i}$. Besides, $P_{cpu}$ is the CPU power of the IoT device on which the module $v_{n,j}$ runs.
\par
The energy consumption for data transmission between the module $v_{n,j}$ and its predecessors $\mathcal{P}(v_{n,j})$ of the application belonging to the $n$th IoT device is calculated as follows:

\begin{eqnarray}
	\label{eq.transmissionEnergyTotal}
	E_{x_{n,j}}^{tra}=\max (\vartheta^{tra}(e_{n,i,j}^{dsize},(h,i),(h^{\prime},i^{\prime}))),\\
	\forall e_{n,i,j}\in \mathcal{E}_n|v_{n,i}\in \mathcal{P}(v_{n,j}), \nonumber\\
	x_{n,i}=(h,i),x_{n,j}=(h^{\prime},i^{\prime})  \nonumber
\end{eqnarray}

\noindent
where, to reduce the length of equations, we consider $H=(e_{n,i,j}^{dsize},(h,i),(h^{\prime},i^{\prime}))$. Similar to response time model, $(h,i)$ and $(h^{\prime},i^{\prime})$ show the specifications of source and destination servers on which modules $v_{n,i}$ and $v_{n,j}$ runs, respectively. The transmission energy consumption between each pair of dependent modules $(\vartheta^{tra}(H))$ is calculated as follows:
\begingroup
\footnotesize
\begin{equation}
	\hspace{-14mm}
	\label{eq.NSEParentTransmission}
	\vartheta^{tra}(H)= \left\{ \begin{tabular}{ll} 
		$(\frac{e_{n,i,j}^{dsize}}{B_{up}}\times P_{t})+(\gamma^{tra}(H^{\prime})\times P_{i}),$&$NSE_{i}(H)=NSE_{1}$\vspace{0.1cm}\\
		$(\frac{e_{n,i,j}^{dsize}}{B_{down}}\times P_{t})+(\gamma^{tra}(H^{\prime})\times P_{i}),$ & $NSE_{i}(H)=NSE_{2}$\vspace{0.2cm}\\
		$\gamma^{tra}(H^{\prime})\times P_{i}$,& $NSE_{i}(H)=NSE_{3}$
	\end{tabular}\right.
\end{equation}
\endgroup


\noindent
where $P_{t}$ presents the transmission power of the IoT device, and the $NSE_{i}$ shows transmission configuration based on $H$. 

\begingroup
\footnotesize
\begin{eqnarray}
	\hspace{-14mm}
	NSE_{i}(H)= \left \{ \begin{array}{llc}
		H^{\prime}=(e_{n,i,j}^{dsize},Par(h,i),(h^{\prime},i^{\prime})),&\text{if }h < h^{\prime}\;\;\&&i=1\\&\;\; h=0, &\\\\
		
		H^{\prime}=(e_{n,i,j}^{dsize},(h,i),Par(h^{\prime},i^{\prime})),&\text{if }h > h^{\prime}\;\;\&&i=2\\&\;\; h^{\prime}=0,& \\\\
		
		H^{\prime}=H,&otherwise,&i=3 
	\end{array}\right.
	\label{NSEcontrol}
\end{eqnarray}
\endgroup

%
%

\noindent
$NSE_1$ states the data flow is starting from an IoT device as the source server to remote servers as destination. Hence, the respective transmission energy consumption is equal to the required time to send the data to the parent server of IoT device multiplied by $P_{t}$, plus the IoT device's idle time (in which the data is transmitted from parent server to the destination) multiplied by $P_i$. Moreover, $NSE_2$ represents the invocation starting from remote servers as the source to the IoT device as the destination. It is important to note that the transmission power of IoT device $P_t$ is active only if one of the modules is assigned to the IoT device and another module run on the remote servers, because we only consider the energy consumption from the IoT device's perspective. In other conditions, the transmission energy consumption is equal to the transmission time $\gamma^{tra}$ (obtained from Eq.~\ref{eq.NSParentTransmission}), in which the IoT device is in idle mode, multiplied by $P_i$ ($NSE_3$).   

\par
The inter-nodal energy consumption $E_{x_{n,j}}^{lat}$ between servers on which module $v_{n,j}$ and its predecessors $\mathcal{P}(v_{n,j})$ are placed is calculated as:

\begin{eqnarray} 
	E_{X_{n,j}}^{lat}=\max  (\vartheta^{lat}((h,i),(h^{\prime},i^{\prime}))), \\
	\forall e_{n,i,j}\in \mathcal{E}_n|v_{n,i}\in \mathcal{P}(v_{n,j}), \nonumber\\
	x_{n,i}=(h,i),x_{n,j}=(h^{\prime},i^{\prime})  \nonumber
\end{eqnarray}

\noindent
where $\vartheta^{lat}$ shows the energy consumption incurred due to inter-nodal delay between source and destination servers on which $v_{n,i}$ and $v_{n,j}$ are placed. This latter is calculated similar to transmission energy consumption based on the $NSE_i(A)$ \cite{xu2019computation,goudarzi2020application}. To reduce the equation size, $((h,i),(h^{\prime},i^{\prime}))=A$.

\begingroup
\begin{equation}
	\label{eq.NSEParentLatency}
	\vartheta^{lat}(A)= \gamma^{lat}(A) \times P_{i}
\end{equation}
\endgroup

\noindent
where the $\gamma^{lat}(A)$ is obtained from Eq.~\ref{eq.NSParentLatency}.
%


\subsubsection{Migration weighted cost model}
We assume that the migration of modules belonging to the $n$th IoT device from current servers to new servers only happens due to the mobility of IoT devices. We consider pre-copy memory migration in which the current servers still running while transferring pre-dump to the new servers \cite{wang2019delay,puliafito2019container}. The goal of migration cost model is to minimize the the downtime plus required cost of executing remaining instructions on the new servers. The migration weighted cost model is defined as:

\begingroup
\footnotesize
\begin{equation}
	\min\limits_{w_{1},w_{2} \in [0,1]} \Psi^{mig}((X_n,X^{\prime}_{n}),t), \quad \forall t \in |SchS_n|, \quad \forall n \in \{1,2,\cdots,N\}
	\label{weightOmegaMigrattion}
\end{equation}
\endgroup

where
\begingroup
\footnotesize
\begin{equation}
	\Psi^{mig}((X_n,X^{\prime}_{n}),t) = w_{1} \times \Gamma^{mig}((X_n,X^{\prime}_{n}),t) + w_{2} \times \Theta^{mig}((X_n,X^{\prime}_{n}),t)
	\label{controlOmegaMigration}
\end{equation}
\endgroup

\begingroup
\footnotesize
\begin{eqnarray}
	\label{eq:migrationOmegaSubjectTo}
	s.t.&&C1:\sum\limits_{t=1}^{|SchS_{n}|} \Psi(X^{\prime}_n,t) \leq \sum\limits_{t=1}^{|SchS_{n}|} \Psi(X_n,t)+\epsilon \;
\end{eqnarray}
\endgroup

\noindent
where $\Gamma^{mig}((X_n,X^{\prime}_{n}),t)$ and $\Theta^{mig}((X_n,X^{\prime}_{n}),t)$ represent the additional time and energy consumption incurred by the migration of modules of $t$th schedule in the downtime (when the service is interrupted). The C1 states the service cost for tasks emitted from modules of $n$th IoT device in the new configuration $X^{\prime}_n$ should be less or roughly the same while considering the previous configuration $X_n$. The $\epsilon$ shows an acceptable additional service cost in the migration. 
Moreover, constraints C1, C2, and C3 from Eq.~\ref{weightOmega} are valid here as well.

\paragraph{\textbf{Migration time model}}
The migration time is considered as the execution time required to finish remaining instructions on the new servers plus the downtime. This latter includes the time for suspending the $Cnts$ in current servers, transmission of the dump and states, and $Cnts$' resuming time on the new servers. Since, in the downtime, a specific amount of dump data and states should also be transferred between servers ($dsize^{mig}$), the migration latency $\gamma_{mig}^{lat}((h,i),(h^{\prime},i^{\prime}))$ and migration transmission time between current and new servers $\gamma_{mig}^{tra}(dsize^{mig},(h,i),(h^{\prime},i^{\prime}))$ to transfer this data are also important \cite{puliafito2019container}. Besides, the $Cnt$s' stopping time plus its resuming time are considered as a constant $I^{mig}$. The migration time is defined as:  

\begin{eqnarray}
	\label{eq.mig}
	\Gamma^{mig}((X_n,X^{\prime}_n),t) =Max(\gamma^{mig}(x_{n,i},x_{n,i}^{\prime})),\\ \forall x_{n,i} \in X_n, \forall x^{\prime}_{n,i}\in X^{\prime}_n|v_{n,i}\in SchS_{n,t},{\tiny } \nonumber \\
	x_{n,i}=(h,i),x^{'}_{n,i}=(h^{\prime},i^{\prime})  \nonumber
\end{eqnarray}
where
\begin{eqnarray}
	\label{eq.mig2}
	\gamma^{mig}(x_{n,i},x_{n,i}^{\prime})=\gamma_{mig}^{lat}((h,i),(h^{\prime},i^{\prime}))+I^{mig} \nonumber\\ +\gamma_{mig}^{tra}(dsize^{mig},(h,i),(h^{\prime},i^{\prime}))+\frac{ e_{n,i,j}^{ins,r}}{cpu({x_{n,i}^{\prime}})}   
\end{eqnarray}

\noindent
where $\gamma^{mig}(x_{n,i},x_{n,i}^{\prime})$ represents the migration cost of module $v_{n,i}$ from its current server $x_{n,i}$ to its new server $x_{n,i}^{\prime}$. The $\gamma_{mig}^{tra}$ and $\gamma_{mig}^{lat}$ are calculated based on \ref{eq.NSParentTransmission} and \ref{eq.NSParentLatency}, respectively. Also, $\frac{e_{n,i,j}^{ins,r}}{cpu({x_{n,i}^{\prime}})}$ shows the execution time of remaining instructions of task $e_{n,i,j}^{ins,r}$ on the new server ($h^{\prime},i^{\prime}$). 

\paragraph{\textbf{Migration energy consumption model}} The additional energy consumption of IoT device, incurred by the migration, depends on the execution of remaining instructions and the downtime.

\begin{eqnarray}
	\label{eq.migEnergy}
	\Theta^{mig}((X_n,X^{\prime}_n),t) =Max(\vartheta^{mig}(x_{n,i},x_{n,i}^{\prime})),\\   \forall x_{n,i}\in X_n, \forall x^{\prime}_{n,i}\in X^{\prime}_n|v_{n,i}\in SchS_{n,t},{\tiny } \nonumber \\
	x_{n,i}=(h,i),x^{'}_{n,i}=(h^{\prime},i^{\prime})  \nonumber
\end{eqnarray}
where
\begin{eqnarray}
	\label{eq.mig2Energy}
	\vartheta^{mig}(x_{n,i},x_{n,i}^{\prime})=\vartheta_{mig}^{lat}((h,i),(h^{\prime},i^{\prime}))+I^{mig} \nonumber\\ +\vartheta_{mig}^{tra}(dsize^{mig},(h,i),(h^{\prime},i^{\prime}))+\vartheta^{exe}_{mig}(x^{\prime}_{n,i})
\end{eqnarray}

\noindent
where $\vartheta^{mig}(x_{n,i},x_{n,i}^{\prime})$ represents the amount of energy consumed by the IoT device in the migration of each module of application from its current server $x_{n,i}$ to its new server $x_{n,i}^{'}$. The $\vartheta_{mig}^{tra}$ and $\vartheta_{mig}^{lat}$ represent the energy consumption incurred due to the transmission and migration latency between current and new servers. They are calculated based on \ref{eq.NSEParentTransmission} and \ref{eq.NSEParentLatency}, respectively. Also, the $\vartheta^{exe}_{mig}(x^{\prime}_{n,j})$ shows the energy consumption required for the execution of remaining instructions of task $e_{n,i,j}^{ins,r}$ on the new server ($h^{\prime},i^{\prime}$).  

\begingroup
\footnotesize
\begin{equation}
	\hspace{-6mm}
	\vartheta^{exe}_{mig}(x^{\prime}_{n,i})= \left\{ \begin{tabular}{ccc} $\frac{e_{n,i,j}^{ins,r}}{cpu({x_{n,i}^{\prime}})}\times P_{cpu}$, &\text{if }$x^{\prime}_{n,i}=(h^{\prime},i^{\prime}) \;\&\; h^{\prime}=0$ \vspace{0.1cm}\\
		$\frac{e_{n,i,j}^{ins,r}}{cpu({x_{n,i}^{\prime}})}\times P_{i}$, &\text{if } $x^{\prime}_{n,i}=(h^{\prime},i^{\prime}) \;\&\; h^{\prime}\neq0$
	\end{tabular}\right.
\end{equation}
\endgroup

\subsection{Optimal Decision Time Complexity}
\label{Appendix:OptimalDecisionTimeComplexity}
We assume $M$ servers exist  in the hierarchical fog/edge computing environment and the maximum number of modules in each IoT application is $K$. Each module of an IoT application can be assigned to one of the $M$ candidate servers at a time. Hence, for an IoT application with $K$ modules, the Time Complexity (TC) of finding the global optimal solution for the application placement and the migration is $O(M^K)$. This cost is prohibitively high and prevents us from obtaining the global optimal solution in real-time \cite{zhang2017towards}. Hence, we propose distributed algorithms to find an acceptable solution in a polynomial time for application placement and migration techniques in hierarchical fog computing environments. 
\section{Proposed Technique}
\label{placement}
\begin{figure}[!t]
	\centering 
	\includegraphics[width=3.3in, height=6.5cm, trim=0.1in 0in 0in 0in]{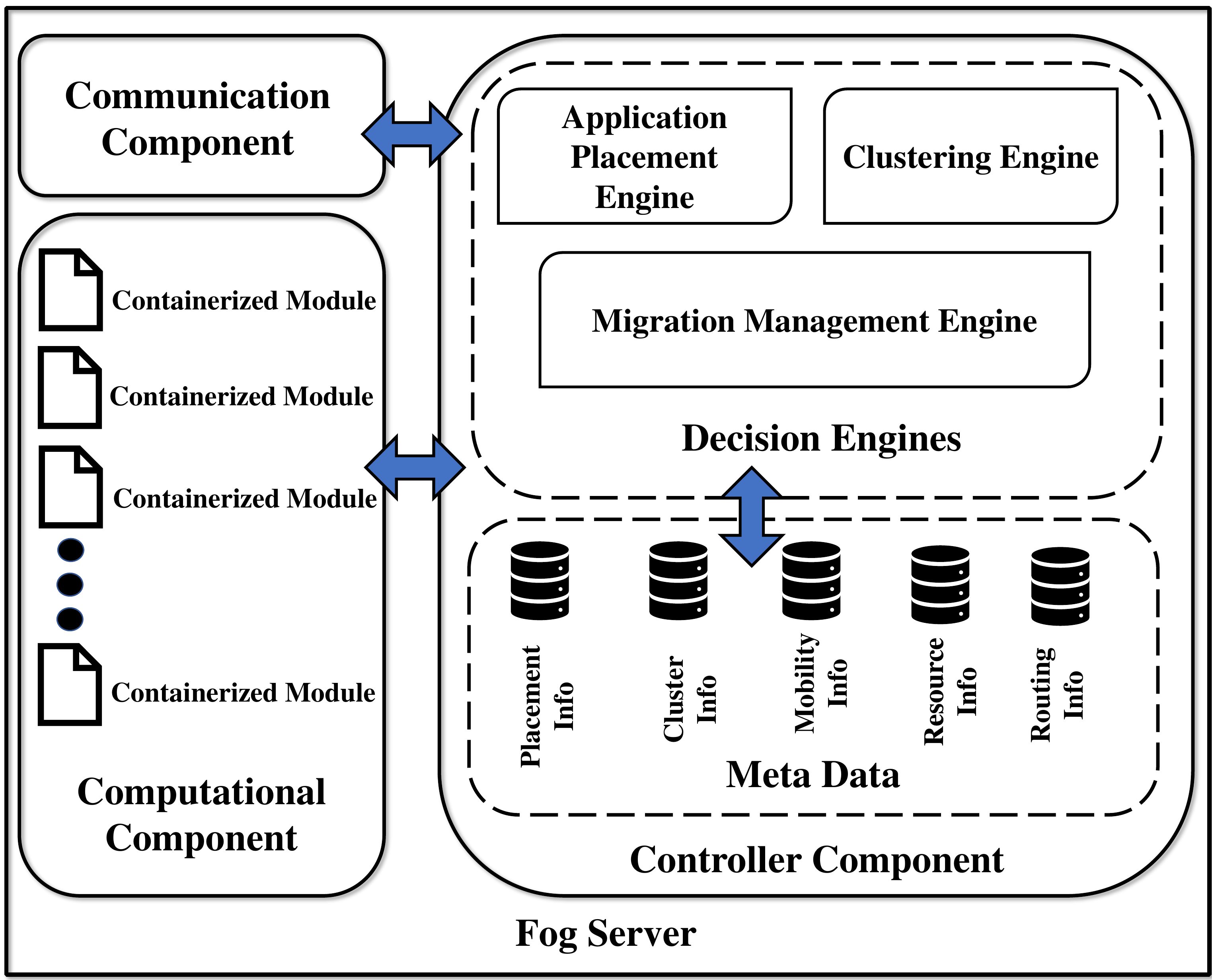}
	\caption{A view of fog server architecture}
	\label{fig:FogServerArchi}
		\vspace{-.6cm}
\end{figure}

In this section, we present a fog server architecture to support distributed application placement, migration management, and clustering (as depicted in Fig.~\ref{fig:FogServerArchi}) by extending the fog server architecture proposed in \cite{mahmud2018latency}. Each FS in \cite{mahmud2018latency} is composed of three main components: controller, computational, and communication. We extend this architecture to support clustering and mobility management of IoT users in a distributed manner.
\par
In our FS architecture, the Controller Component monitors and manages the Communication and Computational Components. It consists of three decision engine blocks and several meta-data blocks to store important information. The \textit{Clustering Engine} is responsible for forming a distributed cluster with its in-range FSs and updating CMs' information in the \textit{Cluster Info} and \textit{Routing Info} meta-data. The \textit{Application Placement Engine} is responsible for placement of IoT applications' modules to minimize the overall cost of running real-time IoT applications. It checks \textit{Cluster Info}, \textit{Resource Info}, and \textit{Routing Info} meta-data for making placement decision, and updates the \textit{Placement Info} and \textit{Resource Info} meta-data blocks to store the configuration of application modules and available resources in this FS, respectively. The \textit{Migration Management Engine} of each FS controls migration process of applications' modules when IoT users move. This module considers all meta-data blocks including the current mobility information of the users (i.,e \textit{Mobility Info}), and decides the migration destination of application modules. Based on its decision, \textit{Placement Info} and \textit{Resource Info} will be updated to store last changes in the configuration of application modules.
\par
The Computational Component provides resources for the execution of application modules that are assigned to this FS based on the container technology. Besides, the Communication Component is responsible for network functionalities such as routing and packet forwarding, just to mention a few \cite{mahmud2018latency}.  

\subsection{Dynamic Distributed Clustering}
Since FSs usually have fewer resources in comparison to CSs, one FS may not be able to provide service for all modules of one application. Moreover, in some scenarios, several IoT devices are connected to the same FS, and hence, the FS may not be able to serve all application modules of different IoT devices due to its limited resources. Thus, other modules of one application should be placed on either CSs or higher-level FSs for the execution. However, in a hierarchical fog computing environment, in which the potential clustering of FSs is considered, application modules can be placed or migrated to other FSs in the same cluster. It can reduce the placement and migration cost of application modules.   
\par  
We consider that FSs belonging to the same hierarchical layer can form a cluster by any in-range FSs at the same hierarchical level and swiftly communicate together using the Constrained Application Protocol (CoAP), Simple Network Management Protocol (SNMP), and so forth. Therefore, the communication delay within a cluster is lower than communication using up-link and down-link \cite{mahmud2018latency}. Besides, in a reliable IoT-enabled system, it is expected that the fog infrastructure providers have applied efficient networking techniques to ensure steady communication among the FSs through less variable inter-nodal latency \cite{mahmud2018latency}. Algorithm~\ref{alg:dynamicClustering} provides an overview of the dynamic distributed clustering technique.
\par
When an FS joins the network, it receives and stores \textit{CandidParent} control messages from FSs residing in the immediate upper layer. The new FS finds coordinates of its position and estimates the average latency to all candidate parents. It selects the FS with the minimum distance as its parent and sends an acknowledgment to it using \textit{ParentSelection} method. Moreover, the new FS broadcasts a \textit{FogJoining} control message, containing its position and coverage range, to its one-hop neighbors (lines 2-7). FSs receiving this message send back a \textit{replyNewFog} control message with their list of active and inactive $Cnts$, positions' specifications, and their coverage range to the new FS. Besides, they update their CM list $List_{cl}$ with specifications of this new FS (lines 8-14). As the new FS receives \textit{replyNewFog} message, it builds its CM list $List_{cl}$ with specifications of FSs residing in the same hierarchical layer. Alongside storing lists of active and inactive $Cnts$ of its CMs, positions, and their coverage range (lines 15-21). This distributed mechanism helps FSs to dynamically update their CM lists when a new FS joins the network. 
\par
We consider that each FS can leave the network in normal conditions (e.g., when the low-level FS is switched off by its user) or due to a failure (such as hardware or software failures). Before an FS leaves the network in normal conditions either permanently or temporarily, we assume that all of its assigned tasks should be finished. Hence, it only needs to send \textit{StartFogLeaving} control message to its CMs to update the $List_{cl}$ of themselves, to its parent server, and to its children to find a new parent (lines 22-25). All FSs that receive \textit{FogLeaving} control message remove all information related to this FS from their entries. Also, the children of the leaving FS that receive this control message call the \textit{ParentSelection} method to update their parent (lines 26-32). In case of a fatal error, in which the leaving FS cannot send a control message to the parent, CMs, and children, its immediate parent runs the \textit{StartFogFailureRecovery} and sends \textit{FogFailureRecovery} control message to its children list $List_{ch}$ so that they can remove entries related to the failed FS (lines 33-39). It is important to note that this latter process takes more time in comparison to the \textit{FogLeaving} process in normal conditions due to the higher latency of uplink and downlink communications. Besides, if any FS children loose their connection to their parent, they can run the \textit{ParentSelection} method to choose a new parent. 
\par
In addition, each FS sends the latest information about its $List_{ch}$ to its parent FS if any changes happen. This helps higher-level FSs update their $\Omega$. 
\setlength{\textfloatsep}{0pt}
\begin{algorithm}[!t]
	\scriptsize
	\caption{Dynamic distributed clustering} \label{alg:dynamicClustering}
	\SetKwData{Left}{left}
	\SetKwData{This}{this}
	\SetKwData{Up}{up}
	\SetKwFunction{Union}{Union}
	\SetKwFunction{FindCompress}{FindCompress}
	\SetKwInOut{Input}{Input}
	\SetKwInOut{Output}{Output}
	\SetKwInOut{Parameter}{Parameter}
	\Input{$RCM$: Received Control Message}
	
	\Switch{$RCM$}{
		\Case{$CandidParent$}{
			ParentSelection()\\
			message.add(getPosition(),coverRange)\\
			message.type($FogJoining$)\\
			Broadcast(message)
		}
		\Case{$FogJoining$}{					
			message.add(getPosition(),coverRange)\\
			message.add(getActiveCnts(),getInactiveCnts())\\
			message.type($ReplyNewFog$)\\
			send($RCM$.getSourceAddr(), message)\\
			$List_{cl}$.update($RCM$.getData())
		}
		\Case{$ReplyNewFog$}{
			$List_{cl}$.update($RCM$.getData())\\
			$MapActiveCnt_{cl}$.put($RCM$.getSourceAddr(),\\message.getListActiveCnts())\\
			$MapInActiveCnt_{cl}$.put($RCM$.getSourceAddr(),\\message.getListInActiveCnts())
		}
		\Case{$StartFogLeaving$}{
			message.type($FogLeaving$)\\
			Broadcast(message)
		}
		\Case{$FogLeaving$}{
			$List_{cl}$.remove($RCM$.getSourceAddr())\\
			$List_{ch}$.remove($RCM$.getSourceAddr())\\
			\If{$RCM$.getSourceAddr() $==$ this.Parent)}
			{	
				
				ParentSelection()\\	
				
			}
			
		}
		\Case{$StartFogFailureRecovery$}{
			\For{$i=1$ to $List_{ch}$.size()}{
				message.type($FogFailureRecovery$)\\
				message.setFailedFog(failedFog.getAddr())\\
				send($List_{ch}$.get(i).getSourceAddr(),message)
			}	
		}
		
		\Case{$FogFailureRecovery$}{
			$List_{cl}$.remove($RCM$.getFailedFogAddr())
		}
	}
	
\end{algorithm}

\subsection{Application Placement}
Due to the time consuming nature of finding the optimal solution (Section~\ref{Appendix:OptimalDecisionTimeComplexity}) for the application placement problem, a \underline{D}istributed \underline{a}pplication \underline{p}lacement \underline{t}echnique (DAPT) is proposed to find a well-suited solution in a distributed manner (Algorithm~\ref{alg:DAPToverview}).  The DAPT starts whenever an application placement request arrives, and the serving FS tries to place application modules on appropriate servers so that real-time tasks, emitted from modules, can be processed with the minimum cost. Considering the weighted cost (Eq.~\ref{weightOmega}), DAPT attempts to place modules of IoT applications in one/several FSs on the lowest-possible layer while considering the potential of clustering. However, if available resources in that/those FSs are not sufficient, it considers upper layer FSs or/and CSs to place the rest of modules. In this way, DAPT reduces the search space of Eq.~\ref{weightOmega} for each FS by only considering itself, its parent FS, and its CMs, and aims at reducing the overall weighted cost. Moreover, a distributed failure recovery method is embedded in DAPT to recover from possible failures.
\begin{algorithm}[!t]
	\scriptsize
	\setlength{\AlCapSkip}{1em}
	\caption{An overview of DAPT} \label{alg:DAPToverview}
	\SetKwData{Left}{left}
	\SetKwData{This}{this}
	\SetKwData{Up}{up}
	\SetKwFunction{Union}{Union}
	\SetKwFunction{FindCompress}{FindCompress}
	\SetKwInOut{Input}{Input}
	\SetKwInOut{Output}{Output}
	\SetKwInOut{Parameter}{Parameter}
	\Input{$\mathcal{G}_{n}$: The DAG of $n$th IoT device,
		$U_{\mathcal{G}_{n}}$: A subset of unassigned modules from $\mathcal{G}_{n}$,	
		$X_{n}$: The configuration of assigned modules,
		$controller_{ID}$: ID of the placement controller
		
	}
	\Output{$X_{n}$}
	$s_{ID}$: this.ID\\
	$List_{cl}$: this.getClusterMembers()\\
	
	\uIf{(controller(n) $||$ ReqFromChild) $\&$ !DAPTFailureRecovery(n)}
	{	
		$List_{cl}^A$=ClusterCheck($List_{cl}$)\\
		$S_{R}$=ReadyServers($List_{cl}^A$,this.parent,$s_{ID}$)\\
		$SchS_{n}$=FindOrder($\mathcal{G}_{n}$)\\	
		$U_{(\mathcal{G}_{n})}$=Sort($U_{(\mathcal{G}_{n})}$, $SchS_{n}$)\\
		\uIf{$S_R - Par(s_ID) \neq \varnothing$}
		{
			\For{$i=1$ to $U_{\mathcal{G}_{n}}$.size()}{
				v=$U_{(\mathcal{G}_{n}),i}$\\
				$ID_{min}$=FindMinCost($S_{R}$,$\mathcal{G}_n$,$X_{n}$,$v$)\\			
				\If{$ID_{min}$ == $s_{ID}$}
				{
					$res_v$=CalService(v)\\
					\eIf{this.$Cnts$.contains(v) $\&$}{
						ScaleCnts(v,$res_v$)\\					
					}
					{
						StartCnt(v)\\
					}
					
					UpdateConfig($X_n$,v,$s_{ID}$)\\
					
				}
				\Else{
					$ReqList$.update($v$,$ID_{min}$)
					
				}
				
				
				
			}
			
			PlaceReqToServers($ReqList$,$\mathcal{G}_{n}$,$X_n$,$S_R$,$TO_{n}$,$SchS_{n}$)


			
			
		}
		\Else{
			PlacePar($\mathcal{G}_{n}$,$U_{\mathcal{G}_{n}}$,$X_n$,$TO_{n}$,$SchS_{n}$)\\
		}
		
	}
	
	\uElseIf{!controller(n) $\&$ !DAPTFailureRecovery(n)}
	{
		\For{$i=1$ to $U_{\mathcal{G}_{n}}$.size()}{
			v=$U_{(\mathcal{G}_{n}),i}$\\
			$res_v$=CalService(v)\\
			\eIf{this.$Cnts$.contains(v) $\&$  $res_v \leq$ this.Resource}{
				ScaleCnts(v,$res_v$)\\
				UpdateConfig($X_n$,v,$s_{ID}$)\\
				NotifyController(v, $s_{ID}$,$controller_{ID}$)\\					
			}
			{
				\eIf{$res_v \leq$ this.Resource}
				{
					StartCnt(v)\\
					UpdateConfig($X_n$,v,$s_{ID}$)\\
					NotifyController(v, $s_{ID}$,$controller_{ID}$)\\
				}
				{
					SendDAPTFailureRecovery(n,v,$controller_{ID}$,$s_{ID}$)\\
				}

			}
			
		}
	}
	\Else{
		DAPTFailureRecovery(n,v,$S_{R}$,$X_n$)\\
	}

	
\end{algorithm}
\par
The immediate FS that receives the placement request from an IoT device is considered as the application placement controller ($controller$) for that IoT device. If the controller is performing the placement of a set of modules or a parent FS receives placement request from its children and the failure recovery mode is not active (lines 3-28), the \textit{ClusterCheck} method returns the list of CMs and their available resources (line 4). Then, the list of ready servers $S_{R}$ containing parent FS, current FS, and available CMs is created (line 5). This list contains all servers that current FS considers for the placement of modules in that hierarchical layer. Next, the \textit{FindOrder} method checks either topological order of modules ($TO_n$) are available or not. If it is not available, it considers the DAG $\mathcal{G}_n$ of $n$th IoT application, and using the Breadth-First-Search (BFS) Algorithm finds topological order of all modules, and creates $SchS_{n}$ (line 6). This latter helps to identify modules that do not have any dependency and can be executed in parallel. Then, $Sort$ method defines priority value for modules that can be executed in parallel (i.e., modules with the same topological order) based on non-increasing order of their rank value (line 7). The rank of each module is defined as:

\begingroup
\footnotesize
\begin{eqnarray}
	\hspace{-11mm}
	\label{eq.upwardRank}
	Rank(v_{n,j})= \left\{ \begin{tabular}{cc} $C^{exe}_{n,j} + \max(C^{tra}_{n,j,z}+Rank(v_{n,z}))$ &\text{if } $v_{n,j} \neq exit$ \vspace{0.1cm}\\	
		$\forall v_{n,z}\in Succ(v_{n,j})$,&\vspace{0.5cm}	\\
		$C^{exe}_{n,j}$, &\text{if } $v_{n,j} = exit$
	\end{tabular}\right.
\end{eqnarray}
\endgroup

\noindent
where $C^{exe}_{n,j}$ shows the average weighted execution cost of module $v_{n,j}$, and $C^{tra}_{n,j,z}$ depicts the transmission cost of module $v_{n,j}$ and $v_{n,z}$, which are calculated as: 

\begin{equation}
	\label{eq.upwardRankExecution}
	C^{exe}_{n,j} = w_1 \times \widetilde{T_{x_{n,j}}^{exe}(S_R)} + w_2 \times \widetilde{E_{x_{n,j}}^{exe}(S_R)}
\end{equation}

\begin{equation}
	\label{eq.upwardRankTransmission}
	C^{tra}_{n,j,z} = w_1 \times \widetilde{\gamma^{tra}_{n,j,z}(S_R)}+ w_2 \times \widetilde{\vartheta^{tra}_{n,j,z}(S_R)}
\end{equation}

\noindent
where $\widetilde{T_{x_{n,j}}^{exe}(S_R)}$ and $\widetilde{E_{x_{n,j}}^{exe}(S_R)}$ show the average execution time and energy consumption of each module considering available servers in the $S_R$. The execution time $T_{x_{n,j}}^{exe}$ and energy consumption $E_{x_{n,j}}^{exe}$ of each module per server are obtained from Eq.~\ref{equation:totalTimeExeModule} and Eq.~\ref{equation:totalEnergyExeModule} respectively. Besides, $\widetilde{\gamma^{tra}_{n,j,z}(S_R)}$ and $\widetilde{\vartheta^{tra}_{n,j,z}(S_R)}$ shows the average transmission time and energy consumption between modules $v_{n,j}$ and $v_{n,z}$ considering available servers in the $S_R$. The transmission time $\gamma^{tra}_{n,j,z}$ and transmission energy consumption $\vartheta^{tra}_{n,j,z}$ between each pair of servers in the $S_R$ can be obtained from Eq.~\ref{eq.NSParentTransmission} and Eq.~\ref{eq.NSEParentTransmission}, respectively. Moreover, $w_1$ and $w_2$ are control parameters to tune the weighted cost. The rank is calculated recursively by traversing the DAG of application, starting from the exit module.
The $Sort$ method can find the critical path of the DAG and gives higher priority to the modules that incur higher execution cost among modules that can be executed in parallel. Hence, the probability of placement of these modules on lower-level FSs increases. This latter is important since the resources of lower-level FSs are limited compared to higher-level FSs, but they can be accessed with less communication cost. Hence, if modules are more communication and latency-sensitive, they can be placed on lower-level FSs with higher priority while if they are computation-intensive modules, that cannot be efficiently executed on the lower-level FSs, they can be forwarded to higher-level FSs with higher priority. If $S_R$ contains any candidate server except its parent, for each module $v$ of $U_{\mathcal{G}_n}$, the \textit{FindMinCost} receives the $S_{R}$, $\mathcal{G}_n$, and configuration $X_n$, as its input and finds the minimum cost for the execution of the module $v$ based on current solution configuration $X_n$ (i.e., based on the assigned servers' configuration to the predecessors of this module). Although in fog computing environments, a large number of FSs are deployed as candidate servers, the DAPT only considers FSs in the $S_R$, to which the serving FS can communicate with the lowest possible transmission and inter-nodal cost. Moreover, we assume that FSs do not have a global view of all FSs in the environment. Therefore, the search space in each hierarchical layer is reduced while the suitable candidate servers for real-time and latency-sensitive IoT applications are kept. After prioritizing modules, the execution cost of each module based on the available servers in $S_R$ is calculated using \textit{FindMinCost} method.   
This method checks the available resources required to run or scale the $Cnts$ to run these modules on available servers. Then, among the servers that meet these requirements, it returns the ID of the selected FS, $ID_{min}$, that can execute module $v$ while minimizing the overall application cost using Eq.\ref{weightOmega} (line 11). If the current FS is selected, and it has active $Cnt$, the \textit{ScaleCnt} method scales the resources so that it can serve this module (line 15). If there is no active $Cnt$ in this FS, it should run a new $Cnt$, which incurs a $Cnt$ startup cost (line 17). The candidate solution configuration $X_n$ is updated accordingly so that the new configuration can be considered for the placement of the rest of the modules (line 19). If the selected FS is among the CMs or parent FS, the module $v$ and its corresponding assigned server are stored in the request list $ReqList$ (line 22) so that it can be forwarded to their destination using the \textit{PlaceReqToServers} (line 25). This method sends modules to assigned serves along with the topological order of this IoT application $TO_n$, schedules $SchS_n$, and current solution configuration $X_n$. Finally, in a case that the $S_R$ is empty, meaning that the current controller does not have any resources and also it does not have any candidate servers with sufficient resources, it sends all modules to the parent FS so that the placement can be started in the higher hierarchical levels by means of the \textit{PlacePar} method (line 27). If the parent FS receives the placement request from its children, it checks the possibility of placement of received modules on its $S_R$. The background reason is if one FS receives some modules for placement from its children FSs, it means that those modules are either more computation-intensive rather than latency/communication-intensive, or the children FSs did not have sufficient resources for these modules. However, if one FS receives a placement request from its CMs, it starts the deployment of modules on the condition that the available resources meet the modules' requirements.

\par
If serving FS is not the controller FS and the failure recovery mode is not active (i.e, the placement request is forwarded to CMs), it iterates over the received modules (i.e., $U_{\mathcal{G}_n}$) and calculates the required amount of resources for each module \textit{CalService(v)}. If it has enough resources, it starts the module, and using \textit{NotifyController} method sends an acknowledgment for the controller FS. However, if due to any problem this FS cannot place this module, it runs \textit{SendDAPTFailureRecovery} method, which sends a failure message to the controller FS so that the controller can make a new decision (lines 29-47).
\par	 
If failure recovery mode is active, it means that one or several servers cannot properly execute assigned modules. Hence, the DAPT algorithm calls \textit{DAPTFailureRecovery} method. This method receives failed modules of $n$th IoT application and finds corresponding FSs from the solution configuration $X_n$. If it has several candidate servers in $S_{R}$, it removes specification of the failed FS from $S_{R}$. Then, it iterates over the rest of available servers to finds FSs for these modules that minimize the execution cost. However, if the current FS only has its parent sever in the $S_{R}$, \textit{DAPTFailureRecovery} sends a control message to activate \textit{DAPTFailureRecovery} method of the parent FS. (line 48). It helps to check the possibility of placement of these modules in higher hierarchical layers.

\subsection{Migration Management Technique (MMT)}
As the user of $n$th IoT device is moving away from its current low-level FS (i.e., its controller FS) to a new low-level FS, the current controller FS should initiate the migration process to find a new controller FS, and migrate the current data and states of running $Cnts$ to new FSs. We suppose IoT devices can detect distributed low-level FSs (eg., using beacons, GPS, etc) and update their list of sensed FSs $List_{SFog}^n$ periodically. Whenever the controller FS realizes that the IoT device $n$ is about to leave (e.g., through the received signal to noise ratio), it receives $List_{SFog}^n$ from the IoT device and initiates the migration process. The goals of the \underline{m}igration \underline{m}anagement \underline{t}echnique (MMT) is to 1) find a new controller FS with the maximum sojourn time for the IoT device and 2) find a set of substitute servers for processing of IoT application's modules while minimizing the migration cost (Eq.~\ref{weightOmegaMigrattion}). The Algorithm \ref{alg:Migration} shows an overview of the distributed migration process.
\par
Whenever a controller FS realizes the $n$th IoT device is about to leave its coverage range, it initiates \textit{MigrationInitiate} to find a new controller FS for the IoT device. The current controller FS receives the list of sensed low-level FSs $List_{SFog}^n$ from $n$th IoT device and removes its $s_{ID}$ from this list so that it cannot be selected as a new controller FS (line 4). The mobility information of each user \textit{mobInfo(n)} contains its average speed and its direction. Moreover, in the clustering technique, each FS learns the position and coverage ranges of its CMs. Considering the aforementioned values, the controller FS can estimate the sojourn time of this IoT device for each CM. The \textit{MobilityAnalyzer} method (line 5) receives \textit{mobInfo(n)} and $List_{SFog}^n$ and checks whether the $List_{SFog}^n$ contains any CMs of the current FS controller. Moreover, it finds specifications of other FSs belonging to $List_{SFog}^n$ through its CMs, if possible. The \textit{MobilityAnalyzer} then creates two separate lists for reachable FSs ($List_{reach}$) and unreachable FSs ($List_{unreach}$) from $List_{SFog}^n$. The former one contains any FSs of $List_{SFog}^n$ which are among CMs of the current controller FS or those that can be accessed through its CMs, while the latter one refers to FSs to which the controller FS does not have access either directly or through its CMs. The \textit{MobilityAnalyzer} method gives higher priority to FSs of $List_{reach}$ because the required information for the new controller to start its procedures can be more efficiently transferred to these FSs compared to those FSs to which it does not have direct access. The MMT considers $resources$ of FSs belonging to $List_{reach}$, and if they have enough resources to serve modules that are currently assigned to the current controller FS, it estimates the sojourn time of $n$th IoT device for those candidate FSs. Then, it returns the ID of the FS with sufficient resources and the maximum estimated sojourn time. 
It is important to note that assigning the controller role to a new FS with maximum sojourn time can reduce the number of possible future migrations, which leads to fewer service interruptions due to migration downtime. On the condition that no FSs of $List_{reach}$ contains enough resources, it returns the ID of FS with the maximum sojourn time. However, if $List_{reach}$ is empty, this method returns the ID of one of the FSs from $List_{unreach}$ randomly. Then, current controller FS sends a \textit{NewControllerReq} message to $dest_{ID}$, containing the DAG of $n$th IoT device application $\mathcal{G}_n$, \textit{mobilityInfo(n)}, and the current configuration of assigned servers $X_n$ (lines 7-9).

\begin{algorithm}[h]
	\scriptsize
	\caption{Migration Management Technique} \label{alg:Migration}
	\SetKwData{Left}{left}
	\SetKwData{This}{this}
	\SetKwData{Up}{up}
	\SetKwFunction{Union}{Union}
	\SetKwFunction{FindCompress}{FindCompress}
	\SetKwInOut{Input}{Input}
	\SetKwInOut{Output}{Output}
	\SetKwInOut{Parameter}{Parameter}
	
	\Input{$RCM$: Received Control Message,
		$\mathcal{G}_{n}$: The DAG of $n$th IoT device,
		$mobInfo(n)$: The mobility data of the IoT device $n$,  
		$X_{n}$: The configuration of assigned modules,
		$controller_{ID}$: ID of the controller,
		$List_{SFog}^n$: Sensed fog devices' List of IoT device $n$
	}
	
	
	\Switch{$RCM$}{
		\Case{$MigrationInitiate$}{
			
			$List_{SFog}^n$=$List_{SFog}^n$.remove($s_{ID}$)\\
			$dest_{Id}$=MobilityAnalyzer(n,$mobInfo$,$List_{cl}$,$List_{SFog}^n$)\\
			message.add($\mathcal{G}_{n}$,$mobInfo(n)$,$X_n$,$TO_{n}$,$SchS_{n}$)\\
			message.type(NewControllerReq)\\
			send($dest_{ID}$,message)\\
			$controller_{pre}(n)$=true\\
			
		}
		\Case{$NewControllerReq$}{
			n=RCM.getIoTDevice\\
			getcontrollerList().add(n)\\
			$X_n$=RCM.getConfig(n)\\
			$ID_{PreCon}$=RCM.getSourceAddr()\\
			$List^{sorted}_{Cnts}$ =SortCntsSize($\mathcal{G}_{n}$, $Cnts^{ram}$) \\
			$MapServer_{pre}$=FindPreServersConfig($X_n$) \\
			\For{$t=1$ to $|SchS_{n}|$}{
				sendMigReqToServers($MapServer_{pre}$,$List^{sorted}_{Cnts}$,$SchS_{n.t}$)\\
				WaitForServersNotifications()\\
			}
			
		}
		\Case{$MigrationReq$}{
			$ReqInfo$=RCM.getInfo()\\
			$Modules$= $ReqInfo$.getModules()\\			
			$S_R$=ReadyServers(this.getCMs(),this.getID(),this.getChildren())\\
			\If{!$S_R$.isEmpty()}{
				\For{$i=1$ to $Modules$.size()}{
					$SortedCostList$=$\varnothing$\\
					\For{$j=1$ to $S_R$.size()}{
						$MigCostTemp$=CalMigCost($Modules_{i}$,$S_{R,j}$)\\
						CostList.update($S_{R,j}$,$MigCostTemp$)\\
					}
					$SortedCostList$=Sort(CostList)\\
					$Server_{ID}$=FindMigrationDestination($SortedCostList$)\\
					sendMigrationDestination($Modules_{i}$,$X_n$,$Server_{ID}$)\\
				}
			}
			\Else{
				SendMigReqToServers(this.Parent(),$ReqInfo$)\\	
			}
			
		}	
		
		\Case{$MigrationDestination$}{
			$v$=RCM.getModule()\\
			$res_v$=calService($v$)\\
			\eIf{$res_v \leq$ this.resources}
			{ 
				sendMigrationStart($v$,$FS_{pre}^{v}$,$FS_{new}^{v}$)\\
				UpdateConfig($X_n$,$v$,$s_{ID}$)\\
				NotifyController($v$,$s_{ID}$,$controller_{ID}$)\\
			}
			{
				SendMMTFailureRecovery(n,v,$controller_{ID}$,$s_{ID}$)\\
			}
			
		}
		\Case{$StartMigration$}{
			
			Migrate($v$,RCM.$FS_{new}^{v}$)\\
			UpdateResoure($v$)\\
			\If{$controller_{pre}(n)$ \& MigrationFinish(n)}
			{
				$controller_{pre}(n)$=false\\
				getControllerList().remove(n)\\
			}
			
		}
		\Case{MMTFailureRecovey}
		{
			MMTFailureRecovery($n$,$v$,$controller_{ID}$,$s_{ID}$)\\
		}
	}
\end{algorithm}


\par 
When an FS receives \textit{NewControllerReq} message, it adds the IoT device $n$ to its \textit{controllerList} to serve this IoT device as its new controller FS (lines 11-12). This new controller FS is responsible for the rest of migration management. It retrieves the current configuration $X_n$ and the previous controller ID, $ID_{PreCon}$, from the received message $RCM$ (lines 13-14). The \textit{SortCntsSize} method descendingly sorts $Cnts$ based on their allocated runtime Ram $Cnts^{ram}$ (line 15). The background reason is the amount of dump and state to be transferred in the downtime is directly related to $Cnts^{ram}$ \cite{puliafito2019container}. The migration of $Cnts$ with larger $Cnts^{ram}$ incurs higher cost in terms of migration time and energy (Eq.~\ref{weightOmegaMigrattion}). Hence, to reduce the total migration cost, MMT gives higher priority to modules with heavier $Cnts^{ram}$ so that the migration decision can be made sooner, and they can be migrated before other modules. Next, \textit{FindPreServersConfig} method retrieves assigned servers' specifications for all application modules and stores them in $MapServer_{pre}$ (line 16). The migration cost (Eq.~\ref{weightOmegaMigrattion}) is defined as the maximum migration cost for each application module while considering $X_n$ and its new configuration $X_n^{\prime}$. The goal is to minimize this migration cost while it is subject to the condition that the new configuration $X_n^{\prime}$ provides better application execution cost or roughly the same with previous configuration $X_n$ (Eq.\ref{eq:migrationOmegaSubjectTo}). So, the MMT retrieves modules of each schedule based on $SchS_n$ and send their corresponding information alongside $MapServer_{pre}$ and $List^{sorted}_{Cnts}$ to \textit{sendMigReqToServers} method. It creates a list of modules based on the hierarchical layer on which modules are previously assigned. Modules of each hierarchical layer are also sorted based on allocated Ram size, obtained from $List^{sorted}_{Cnts}$. This method sends $MigrationReq$ messages alongside respective modules' information to FSs that are responsible for making the migration decision. As MMT acts in a distributed manner and FSs at each layer only has information about their parent, children, and CMs, migration decisions for modules of each layer are made by the new controller, its parent, or ancestors in the hierarchy. To illustrate, considering Fig.~\ref{fig:systemmodel}, we assume an IoT application has three modules in one of its schedules and two of them were previously assigned on FS (1,3) (prior controller), and one on FS (2,1). If we assume that the new controller is FS (1,4), it makes migration decision for modules that previously assigned on FS (1,3) while $par(1,4)$ (i.e., FS (2,3)) makes migration decision for the module that previously assigned on FS (2,1). After sending migration requests $migrationReq$, FS (1,4) waits to receive notifications and new configuration of modules for that schedule and then iterates over next schedules (lines 17-20).
\par
When an FS receives \textit{MigrationReq} message, the FS retrieves the information and forwarded modules from the received message (lines 23-24). Then, the list of ready servers $S_R$ is created based on CMs, and children. If the $S_R$ does not contain any available servers, all the modules are forwarded to the parent FS for making migration decision (line 39), while if it contains servers, it tries to minimize the migration cost based on the specification of available servers (line 26-37). This FS considers a list of $modules$, sorted descendingly based on $Cnts^{ram}$, for making migration decision. Hence, the migration of modules that incur higher migration costs in each schedule is performed with higher priority, leading to less overall migration costs in that schedule. Then, for each selected module, the migration cost is estimated and stored in the $CostList$ (line 29-32). The \textit{Sort} method sorts the migration costs ascendingly so that servers with lower migration cost receives higher priority (line 33). Then, the $FindMigrationDestination$ method selects a new server for the module, considering $SortedCostList$, which minimizes the migration cost while it does not negatively affect the application's running cost. Hence, this method iterates over $SortedCostList$, sorted ascendingly based on the migration costs, and selects the server that satisfies the Eq.~\ref{eq:migrationOmegaSubjectTo} (line 34). Finally, the \textit{sendMigrationDestination} method sends a $MigrationDestination$ message to the selected FS to check its resources and start the migration of the respective module. 
\par
The FS receiving \textit{MigrationDestination} checks whether it has enough resources to serve the module $v$ or not (lines 42-44). If this FS can serve the module $v$, it sends a \textit{StartMigration} message to the $FS^{v}_{pre}$ so that it can start the migration. Then, it updates the $X_n$ with its $s_{ID}$ and notifies the controller (lines 45-468). If it cannot serve this module due to any reason, it runs the \textit{SendMMTFailureRecovery} method to send a failure message to the controller FS (lines 49-51).
\par
The \textit{MMTFailureRecovey} is working as the same as \textit{DAPTFailureRecovey}. The only difference is that the migration cost in the MMT is obtained from Eq.~\ref{weightOmegaMigrattion} (lines 59-61).

\par                          
Whenever an FS receives a \textit{StartMigration} message, it starts the migration and then frees the previously assigned resources (lines 53-55). Moreover, if the FS was previously the controller for the $n$th IoT device, and it finishes the migration of all assigned modules belonging to that IoT device, the FS removes the $n$th IoT device from its \textit{controllerList} (lines 56-59).  

\subsection{Complexity Analysis}
The Time Complexity (TC) of the clustering phase (Algorithm \ref{alg:dynamicClustering}) depends on the size of $List_{cl}$ and $List_{ch}$, and candidate parents in the immediate upper level for each FS. In the worst-case scenario, if we assume all FSs reside in one cluster and/or they have only one parent. Hence, the TC of \textit{remove} method belonging to the \textit{FogLeaving} and \textit{FogFailureRecovery} is $O(F)$, and the TC of the \textit{StartFogFailureRecovery} is $O(F)$. Moreover, the TC of \textit{ParentSelection} method of \textit{CandidParent} is $O(F)$ in the worst-case scenario if we assume one FS has $F-1$ candidate parent. Hence, the TC of the clustering step in the worst-case scenario is $O(F)$. Moreover, in the best-case scenario, the number of FSs in $List_{cl}$ and/or the size of the $List_{ch}$ is one, and the TC of the best-case is $O(1)$.
\par
To find the TC of DAPT (Algorithm \ref{alg:DAPToverview}), we suppose that the size of the largest IoT application is $K$. So, in the worst-case scenario, the size of $U_{\mathcal{G}_n}$ is $K$. The \textit{FindOrder} method finds the topological order of the DAG using BFS algorithm with the TC of $O(K+|\mathcal{E}|)$, in which $|\mathcal{E}|$ represents the number of data flows. In the dense DAG, the $|\mathcal{E}|$ is of $O(K^2)$. Moreover, the TC of \textit{Sort} Algorithm is $O(FK^2)$ in the worst-case scenario. In the worst-case scenario, all FSs reside in one cluster and have enough resources for any requests. Hence, the worst-case TCs of \textit{ClusterCheck}, \textit{ReadyServers}, \textit{FindMinCost}, and \textit{DAPTFailureRecovery} are of $O(F)$, $O(F)$, $O(FK)$, and $O(FK)$, respectively. Hence, the worst-case TC of DAPT Algorithm is $O(FK^2+FK)$. In the best-case scenario, the DAG of the application can be sparse so that the TC of \textit{FindOrder} and \textit{Sort} algorithms become $O(K)$ and $O(1)$, respectively. Moreover, in the best-case scenario, the number of available servers in one cluster is one, and hence, TCs of \textit{ClusterCheck}, \textit{ReadyServers}, \textit{FindMinCost}, and \textit{DAPTFailureRecovery} are of $O(1)$, $O(1)$, $O(K)$, and $O(K)$, respectively. So, TC of DAPT in the best-case scenario is $O(K)$.  
\par
The TC of the \textit{MigrationInitiate} from Algorithm \ref{alg:Migration} depends on the TC of \textit{MobilityAnalyzer}. In the worst-case scenario, all the FSs reside in one cluster and the IoT device can sense all of them. So, the size of the list of sensed FSs $List^{n}_{SFog}$ is equal to $F$. Hence, in the worst-case, the TC of creating $List_{reach}$ and $List_{unreach}$ is of $O(F^2)$ while in the best-case scenario, it is of $O(F)$ when there is only one FS in the cluster. Moreover, the worst-case TC of finding maximum sojourn time is $O(F)$. So, the TC of \textit{MigrationInitiate} in the worst-case is $O(F^2)$ while in the best-case, it is of $O(F)$. The TC of the \textit{NewControllerReq} in the worst-case is $O(KLogK+FK)$ while TC of \textit{NewControllerReq} in the best-case scenario is $O(KLogK)$ when there is only one FS in each cluster. The $TC$ of \textit{MigrationReq} in the worst-case scenario depends on the TCs of $CalMigCost$ and $Sort$ which are $O(FK)$ and $O(FKLogF)$ while in the best-case scenario they are O(K). The TC of \textit{MigrationDestination} depends on the TC of $MMTFailureRecovery^{mig}$ and is of $O(F)$ at the worst-case and $O(1)$ in the best-case scenario. Therefore, the TC of the MMT in the worst-case scenario is $O(F^2+FK+KLogK+FKLogF)$ while in the best-case scenario is $O(KLogK)$.
\par
Considering TCs of all methods, the TC of our technique in the worst-case scenario is $O(F^2+FK^2+FKLogF)$ while in the best-case scenario, it is $O(F+KLogK)$.

\section{Performance Evaluation}
\label{evaluation}
In this section, the system setup and parameters, and detailed performance analysis of our technique, in comparison to its counterparts, are provided.

\subsection{System Setup and Parameters}
We extended the iFogSim simulator \cite{gupta2017ifogsim} for the implementation and evaluation of distributed mobility management, clustering, and failure recovery techniques. We used DAGs of two real-time applications, namely the Electroencephalography tractor beam game (EEGTBG) \cite{gupta2017ifogsim,bittencourt2017mobility} and ECG Monitoring for Health-care applications (ECGMH) \cite{pallewatta2019microservices} to create our DAGs. Both applications consist of a sensor and display modules that are placed in the IoT device (e.g., smartphone, wearable devices, etc). Other modules can be placed either on distributed FSs or CSs based on the distributed application placement decisions and/or the migration technique. Data transmission intervals for ECG and EEG sensors are 10ms and 15ms, respectively \cite{mahmud2018latency,gupta2017ifogsim}. Besides, we assume the amount of RAM allocated to each container at the runtime for state size is randomly selected from 50-75 MBytes \cite{puliafito2019container}.
The total amount of data to be transferred in the downtime (i.e., $dsize^{mig})$ is just a few MBytes \cite{puliafito2019container}, which is randomly selected from 5-10\% of each container's allocated RAM in the runtime.      
\par  
We simulate a 2km $\times$ 1km area, in which the coverage range of FSs situated in the first and second layers is assumed to be 200m and 400m, respectively. The system consists of one layer of IoT devices, three layers of heterogeneous FSs, and a  layer \cite{mahmud2018latency,taneja2017resource,pallewatta2019microservices}. The IoT device layer consists of 80 IoT devices, while the number of FSs in level 1, level 2, and level 3 are 30, 5, and 1, respectively. The computing power (CPU) of IoT devices is considered as 500 MIPS \cite{xu2019computation}, while the computing power of level 1 FSs is randomly selected from [3000-4000] MIPS \cite{xu2019computation,bittencourt2017mobility}. Besides, the total computing power of level 2 FSs, level 3 FSs, and CS are considered as 8000 MIPS, 10000 MIPS, and 80000 MIPS, respectively \cite{taneja2017resource,pallewatta2019microservices}. Besides, the latencies between IoT devices to level 1 FSs, level 1 FSs to level 2 FSs, level 2 FSs to level 3 FSs, and level 3 FSs to cloud servers are 5ms, 25ms, 50ms, and 150ms, respectively \cite{mahmud2018latency,pallewatta2019microservices,taneja2017resource}. The upstream and downstream network capacity of IoT devices are 100 Mbps and 200 Mbps, respectively. The upstream, downstream, and clusterlink network capacity for FSs and the CSs are also considered to be 10 Gbps \cite{taneja2017resource,pallewatta2019microservices}.  Moreover, clusters can be formed among the level 1 and level 2 FSs with their in-range FSs of the same hierarchical layer. The communication latency among the FSs residing in level 1 clusters and FSs residing in level 2 clusters are [3-5] ms and [20-25] ms, respectively \cite{mahmud2018latency,pallewatta2019microservices}. The processing power consumption, idle power consumption, and transmission power consumption of IoT devices are 0.9W, 0.3W, and 1.3W, respectively \cite{kumar2010cloud,goudarzi2020application}. User trajectories are generated by a variation of the random walk mobility model \cite{sun2017emm,wang2019dynamic}, in which each user selects a direction, chooses a destination anywhere toward that direction, and moves towards it with a uniformly random speed. The user arriving at the destination can choose a new random direction. 

\begin{table}[!ht]
	\caption{Evaluation Parameters}
	\centering
	\label{tab:parameterValues1}
	\footnotesize
	\renewcommand{\arraystretch}{1.2}
	\begin{tabular}{ll}
		\hline
		\multicolumn{1}{l}{\textbf{Parameter }}                                                                                          & \multicolumn{1}{l}{\textbf{Value}}                         \\ \hline
		Simulation Time                                                                                       & 100,200,300,400 (S)                  \\ \hline
		Area                                                                                                  & 2km $\times$ 1km \\ \hline
		Users' Speed                                                                                          & {[}0.5-4{]} m/s             \\ \hline
		\multicolumn{2}{c}{\textbf{Latency (ms)}}                                                                                                   \\ \hline
		\begin{tabular}[c]{@{}c@{}}ECG Sensor Data Transmission Interval\end{tabular}                       & 10                            \\ \hline
		\begin{tabular}[c]{@{}c@{}}EEG Sensor Data Transmission Interval\end{tabular}                       & 15                            \\ \hline
		
		\begin{tabular}[c]{@{}c@{}}ECG and EEG Sensor $\leftrightarrow$  IoT Device\end{tabular} & 2                             \\ \hline
		IoT Device $\leftrightarrow$ Level 1 FS                                           & 5                             \\ \hline
		Level 1 FS $\leftrightarrow$ Level 2 FS                                                  & 25                            \\ \hline
		Level 2 FS $\leftrightarrow$ Level 3 FS                                                  & 50                            \\ \hline
		Level 3 FS $\leftrightarrow$ Cloud                                                       & 150                           \\ \hline
		L1 Clusters                                                                                           &         [3-5]            \\ \hline
		L2 Clusters                                                                                           &          [20-25]        \\ \hline                                                                   
	\end{tabular}
\end{table}


\subsection{Performance Study}
We conducted seven experiments evaluating system size analysis, average execution cost of tasks, cumulative migration cost, the total number of migrations, Total number of Interrupted Tasks (TIT) due to the migration, Failure recovery analysis, and optimality analysis. In the experiments, to obtain the weighted cost of placement and migration, the $w_1$ and $w_2$ are set to 0.5. To analyze the efficiency of our technique, we extended two other counterparts in the dependent category of fog computing proposals as follows:

\begin{itemize}
	
	\item \emph{MAAS}: This is the extended version of the technique called Mobility-Aware Application Scheduling (MAAS) \cite{bittencourt2017mobility} working based on edgeward-placement technique. The main concern of this edge-centric technique is to place dependent modules of IoT applications on remote servers based on their pre-known mobility pattern (i.e., source, destination, and the potential paths between them are known in advance) of users. In MAAS, if an FS cannot place modules on itself, the modules should be forwarded to the parent server for placement. We extended this technique to support the migration as the users move among remote servers in the runtime while considering the destination and potential paths are not priori-known. 
	
	\item \emph{Urmila}: This is the extended version of Ubiquitous Resource Management for Interference and Latency-Aware services (Urmila) \cite{shckhar2019urmila} which proposes a mobility-aware technique for placement of dependent modules of IoT applications while mobility pattern of users are priori-known. In this technique, the central controller is placed in the highest level FS, and makes placement decisions for IoT applications consisting of dependent modules. We extended this technique so that the central controller helps remote servers to migrate dependent modules of applications as the IoT users move.
\end{itemize}

\subsubsection{System size analysis}

\begin{figure}[!t]
	\centering 
	\includegraphics[width=\linewidth, height=3cm]{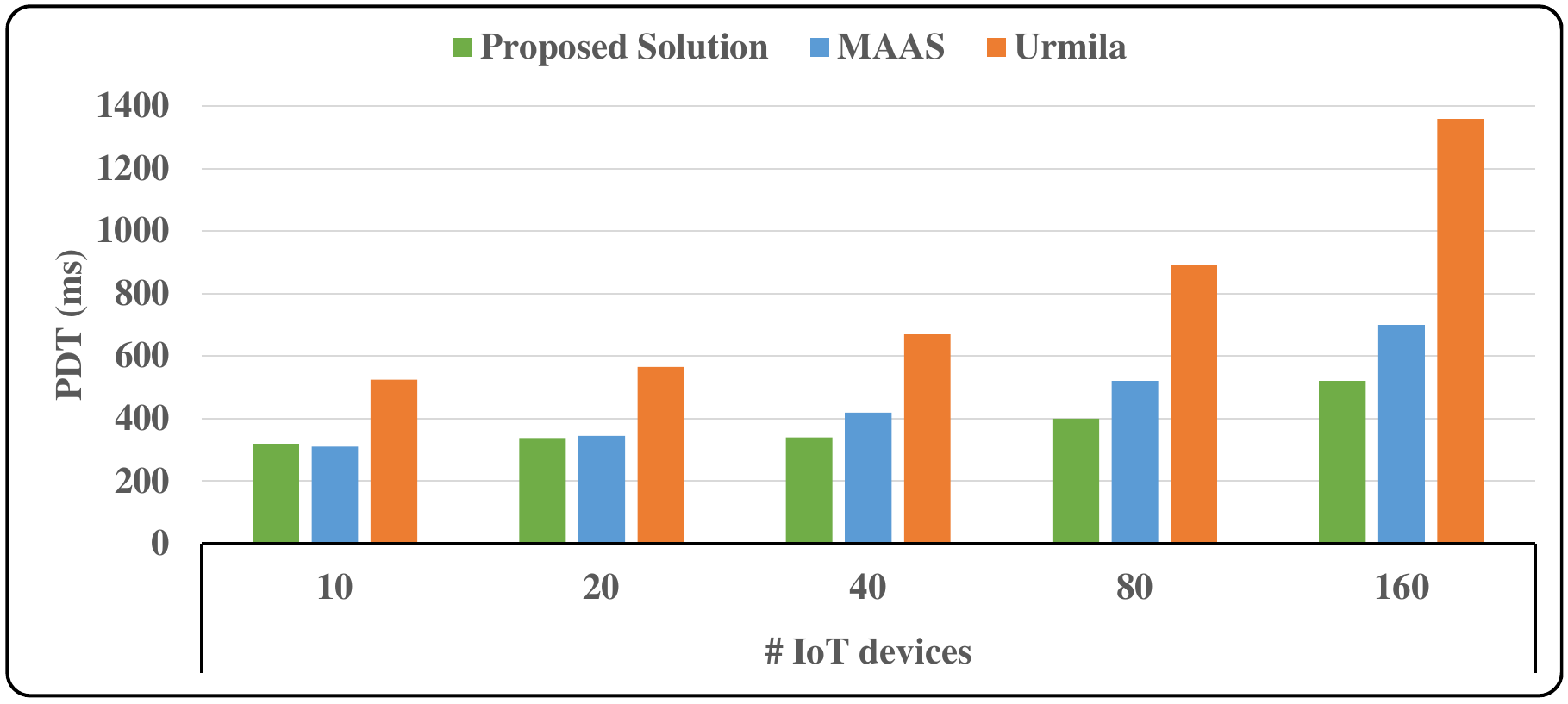}
	\caption{Placement Deployment Time (PDT)}
	\label{fig:PDT}
\end{figure}

In this experiment, we study the effect of number of IoT devices on the Placement Deployment Time (PDT). The PDT shows the period between the start of sending placement requests from IoT devices up to the time the deployment of application modules of IoT devices on FSs are finished. Obviously, the PDT includes the decision time in which FSs make placement decisions and the container startup cost on the servers. Regardless of the quality of solutions that each technique provides, the PDT helps to understand how long the IoT devices should wait until the service can start. In this experiment, the number of IoT devices is increased from 10 to 160 by multiplication of two. Although the number of IoT devices increases in this experiment, we fixed the number of FSs so that we can analyze how different techniques work when the number of placement requests increases significantly. Besides, it is clear that our technique, due to its distributed manner, can easily manage the increased number of placement requests when the number of FSs increases.
\begin{figure*}[!t]
	\begin{subfigure}{.325\textwidth}
		\centering
		\includegraphics[width=\linewidth,height=3cm]{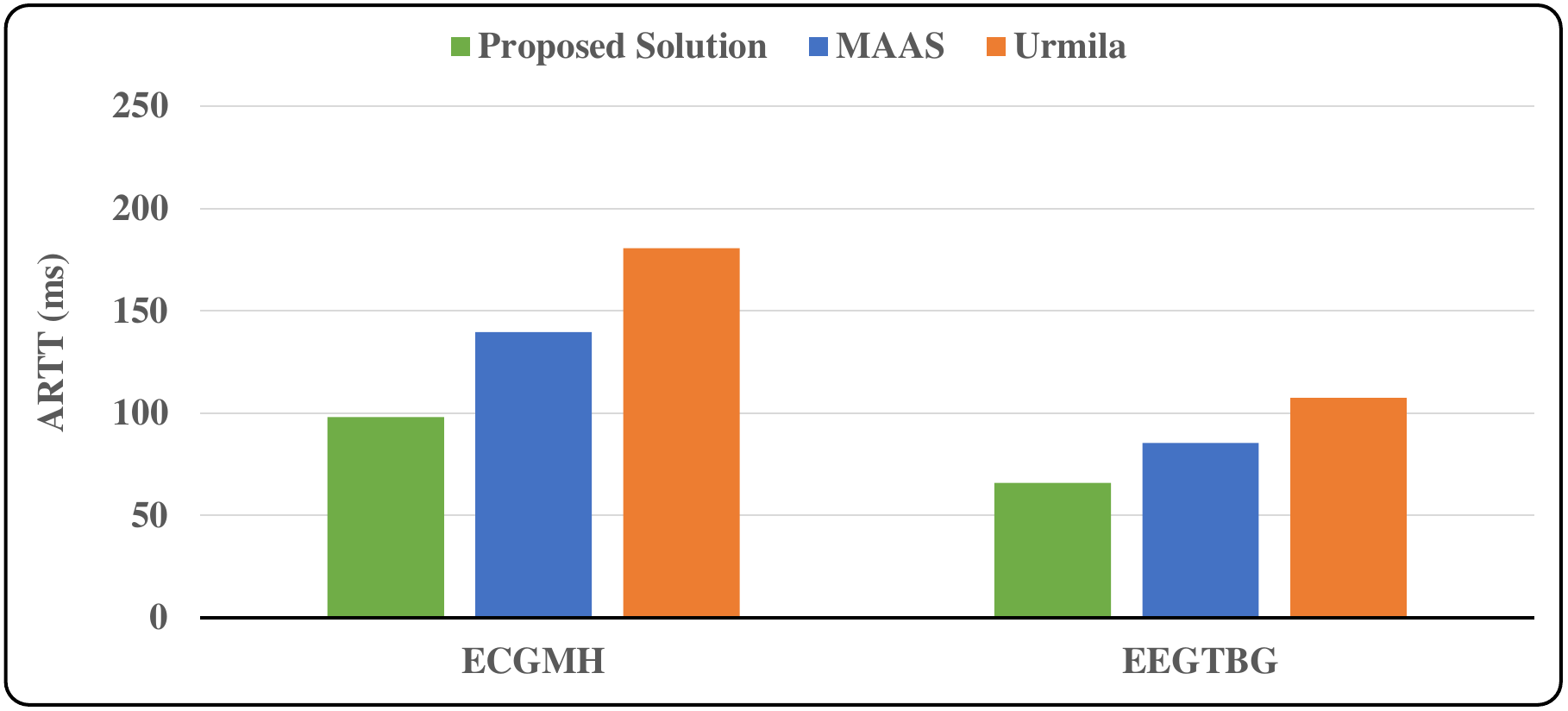}
		\captionsetup{justification=centering}
		\caption{Average Response Time of Tasks (ARTT)}
		\label{fig:AverageLoopCost:sub1}
	\end{subfigure}%
	\hspace{0.1cm}
	\begin{subfigure}{.325\textwidth}
		\vspace{0.35cm}
		\centering
		\includegraphics[width=\linewidth,height=3cm]{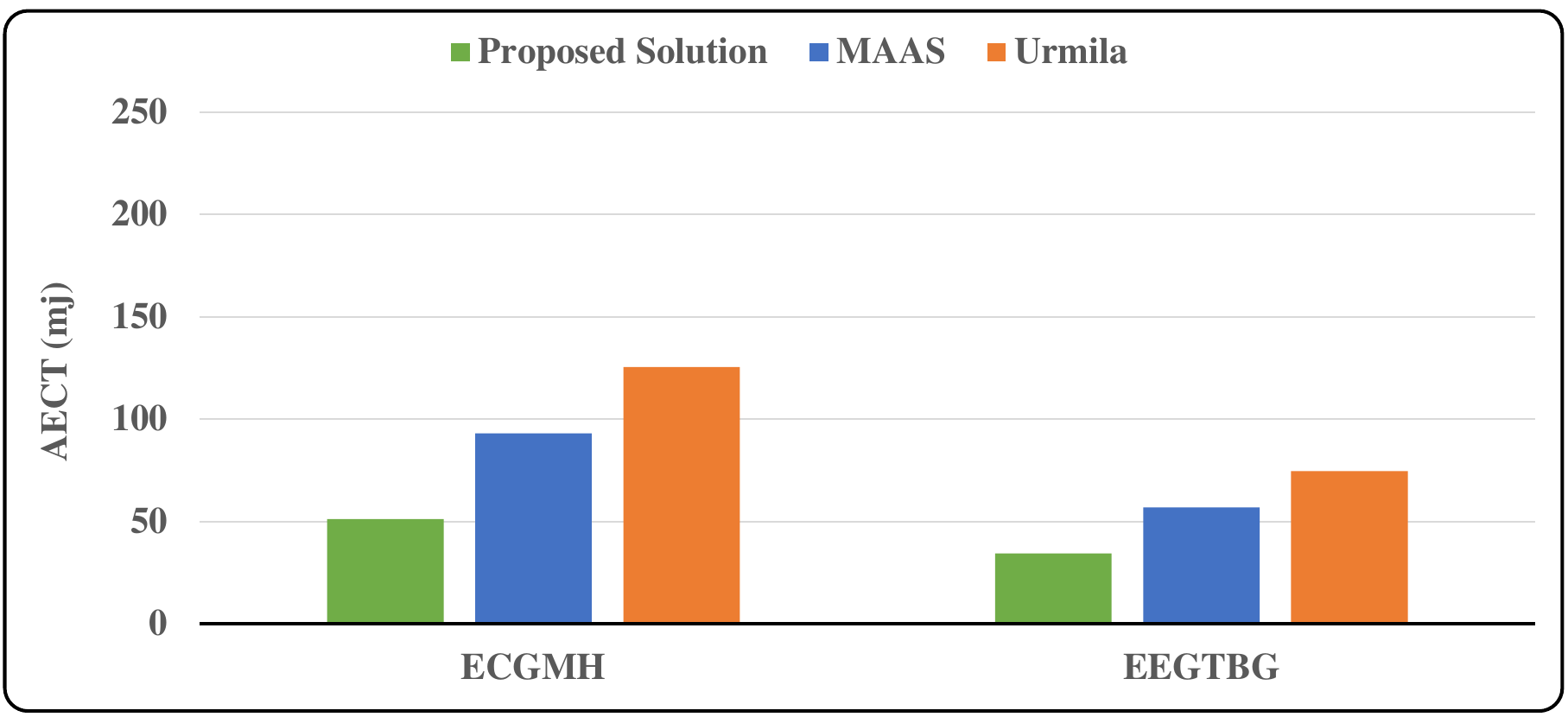}
		\captionsetup{justification=centering}
		\caption{Average Energy Consumption of Tasks (AECT)}
		\label{fig:AverageLoopCost:sub2}
	\end{subfigure}
	\hspace{0.25mm}
	\begin{subfigure}{.325\textwidth}
		\centering
		\includegraphics[width=\linewidth,height=3cm]{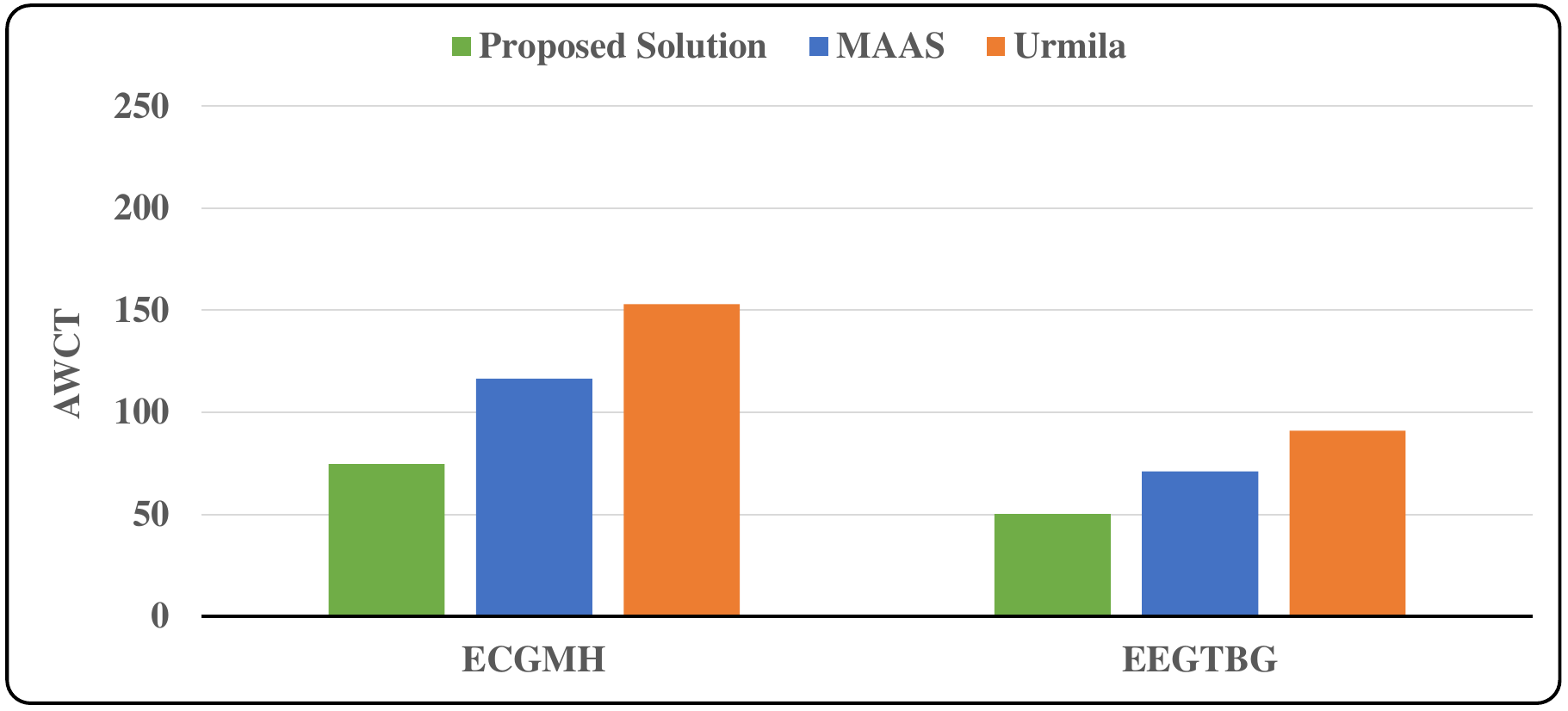}
		\captionsetup{justification=centering}
		\caption{Average Weighted Cost of Tasks (AWCT)}
		\label{fig:AverageLoopCost:sub3}
	\end{subfigure}
\caption{Average execution cost of tasks}
\label{fig:AverageLoopCost}
\end{figure*}  
\par
In Fig.~\ref{fig:PDT}, the PDTs of our proposed solution and MAAS are significantly lower than Urmila, specifically in a larger number of IoT devices. This latter is mainly because our solution and MAAS use a distributed placement engine while Urmila uses a centralized approach. When the placement decision engine receives incoming placement requests, it should make placement decisions and then manage the deployments of application modules in different servers according to solutions' configuration. In Urmila, all of the placement requests should be forwarded to the centralized entity, meaning that the number of arriving placement requests in the decision engine is larger than the distributed placement techniques. Hence, the processing of these requests on the centralized controller takes more time compared to the distributed placement engines, especially when the number of IoT devices increases. Moreover, our solution outperforms the MAAS since it tries to place more application modules in the lowest hierarchical layer, compared to MAAS, which incurs less deployment time. 

\subsubsection{Average execution cost of tasks}

This experiment shows the average execution cost of tasks emitted from a sensor module until they arrive at actuator in 400 seconds of simulation.
\par
As it can be seen from Fig~\ref{fig:AverageLoopCost}, our proposed solution outperforms the MAAS and Urmila in terms of Average Response Time of Tasks (ARTT), Average Energy Consumption of Tasks (AECT), and Average Weighted Cost of Tasks (AWCT). In the MAAS, each FS, from the lowest to the highest hierarchical level, attempts to place modules on itself or forwards them to its parent server for the placement or handling of the migration process. Therefore, it does not consider other potential servers at the same hierarchical level, which incurs higher transmission and inter-nodal costs. The pure Urmila, on the other hand, does not migrate the application modules to servers that are closer to the moving IoT devices, and hence, the average execution cost of tasks, emitted from IoT devices, increases significantly. In our distributed technique, however, each FS considers potential servers at the same hierarchical level (for placement and migration) if those servers are among its CMs. In this way, we decrease the large search space of centralized techniques, while we use the benefits that servers at the same hierarchical level can provide. Also, since modules with higher costs have higher placement priority, the possibility of their placement on more suitable servers are higher compared to other modules. This latter leads to better placement decisions that minimize the cost of executing tasks. It is important to note that the average execution cost of the EEGTBG is lower than the ECGMH. It is because tasks' instruction number in the EEGTBG is lower than of ECGMH ones.

\subsubsection{Total number of migrations}
\begin{figure}[!t]
	\centering 
	\includegraphics[width=\linewidth,height=3cm]{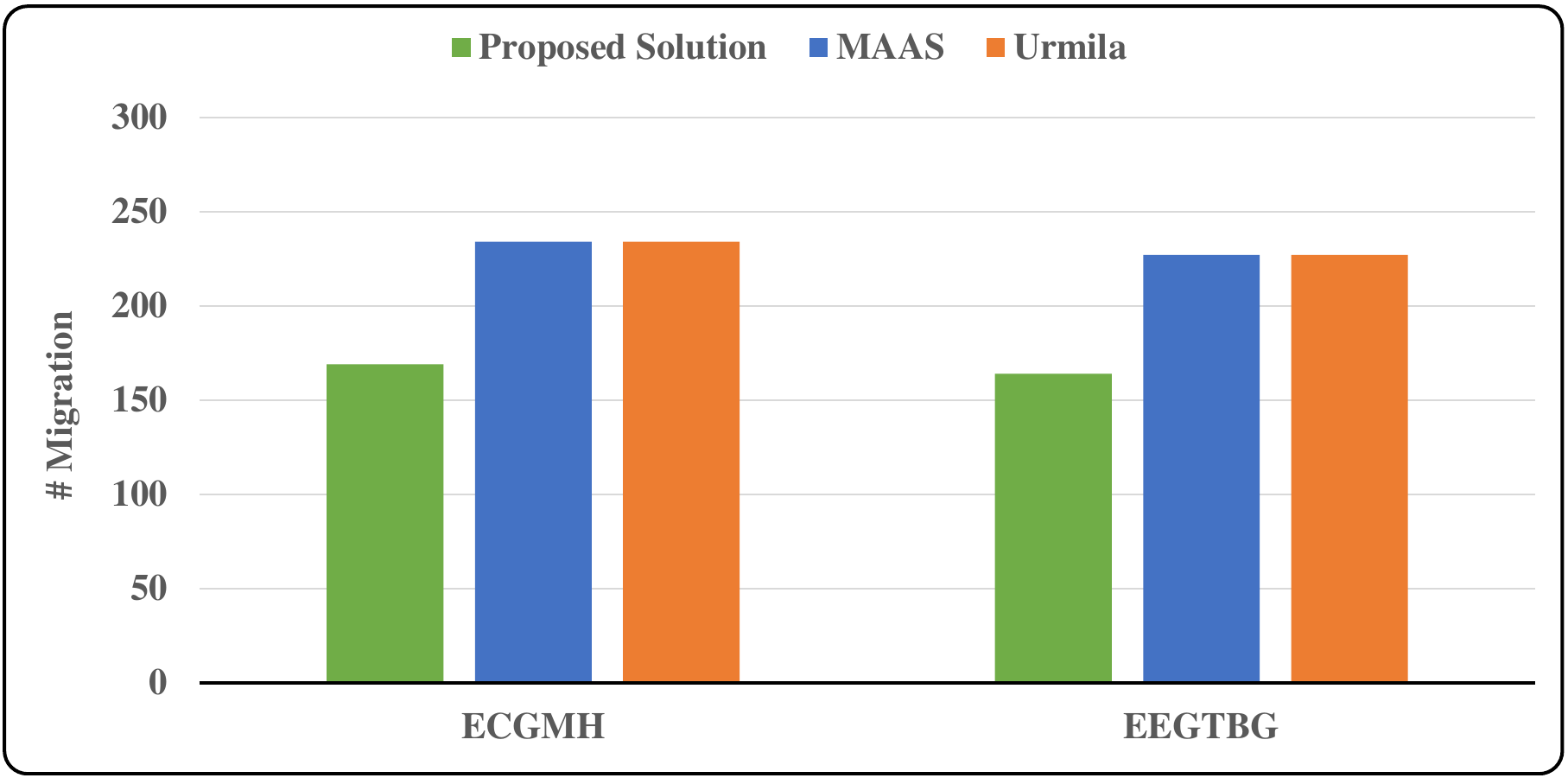}
	\caption{Total number of migrations}
	\label{fig:TotNumMig}
	
\end{figure}
This experiment studies the total number of migrations that occurred during 400 seconds due to the IoT users' movement.
\par   
It can be seen from Fig.~\ref{fig:TotNumMig} that our technique leads to a smaller number of migrations in comparison to its counterparts. This is because our solution considers the current mobility information of IoT devices such as current speed and direction. Since the controller FS has coordinates of its CMs and current mobility information of leaving IoT devices (e.g., their average speed and their direction while in the range of the current controller FS), the serving FS can estimate a sojourn time for all candidate remote servers for the migration. Hence, by the migration of modules to the remote server with the highest sojourn time (in case sufficient resources are available), the number of possible migrations decreases. The extended MAAS and Urmila only try to reduce the migration cost by migrating modules to new remote servers, while they do not consider current mobility information of IoT devices and their sojourn time in remote servers. Hence, they may select remote servers in which the IoT devices stay only for a short period.

\subsubsection{Cumulative migration cost}
\begin{figure*}[!ht]
	\begin{subfigure}{.325\textwidth}
		\centering
		\includegraphics[width=\linewidth,height=3cm]{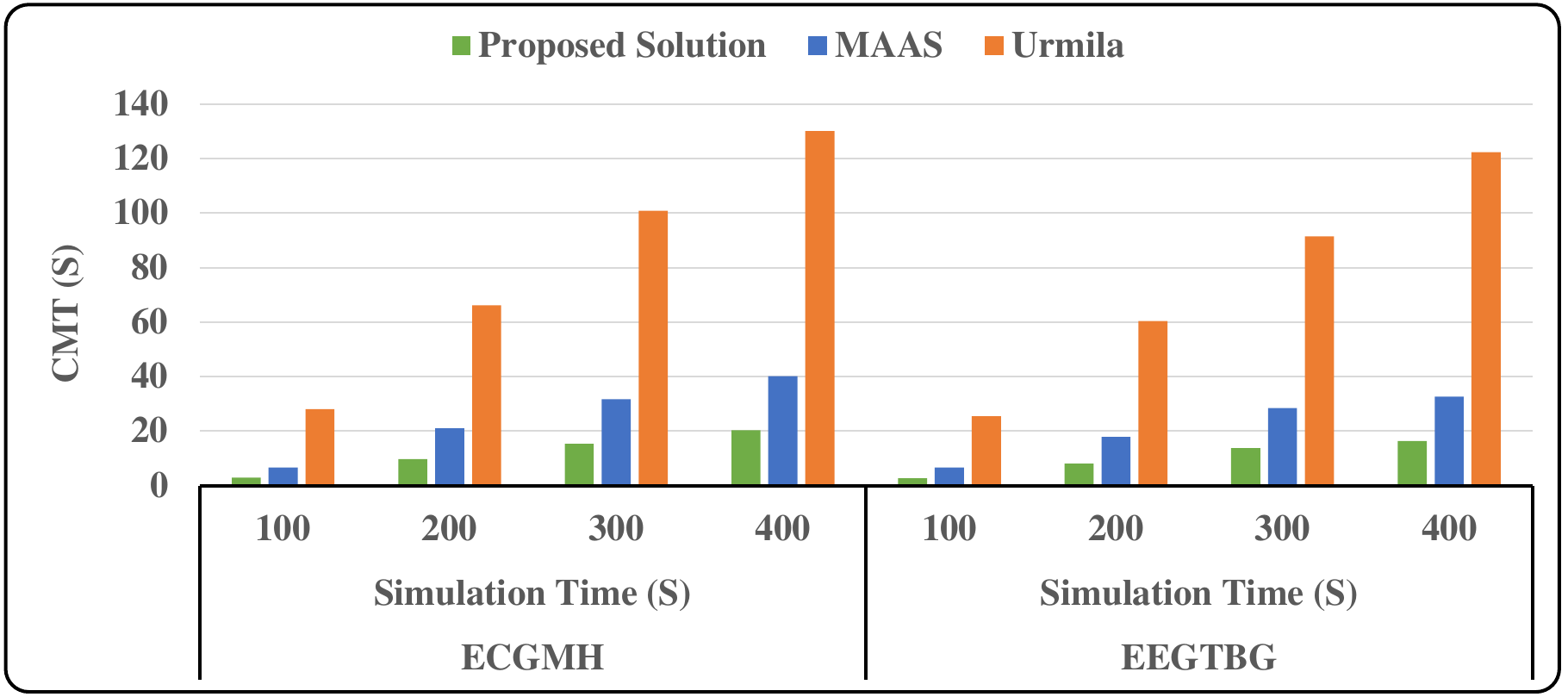}
		\captionsetup{justification=centering}
		\caption{Cumulative Migration Time \\(CMT)}
		\label{fig:CMC:sub1}
	\end{subfigure}%
	\hspace{0.1cm}
	\begin{subfigure}{.325\textwidth}
		\centering
		\includegraphics[width=\linewidth,height=3cm]{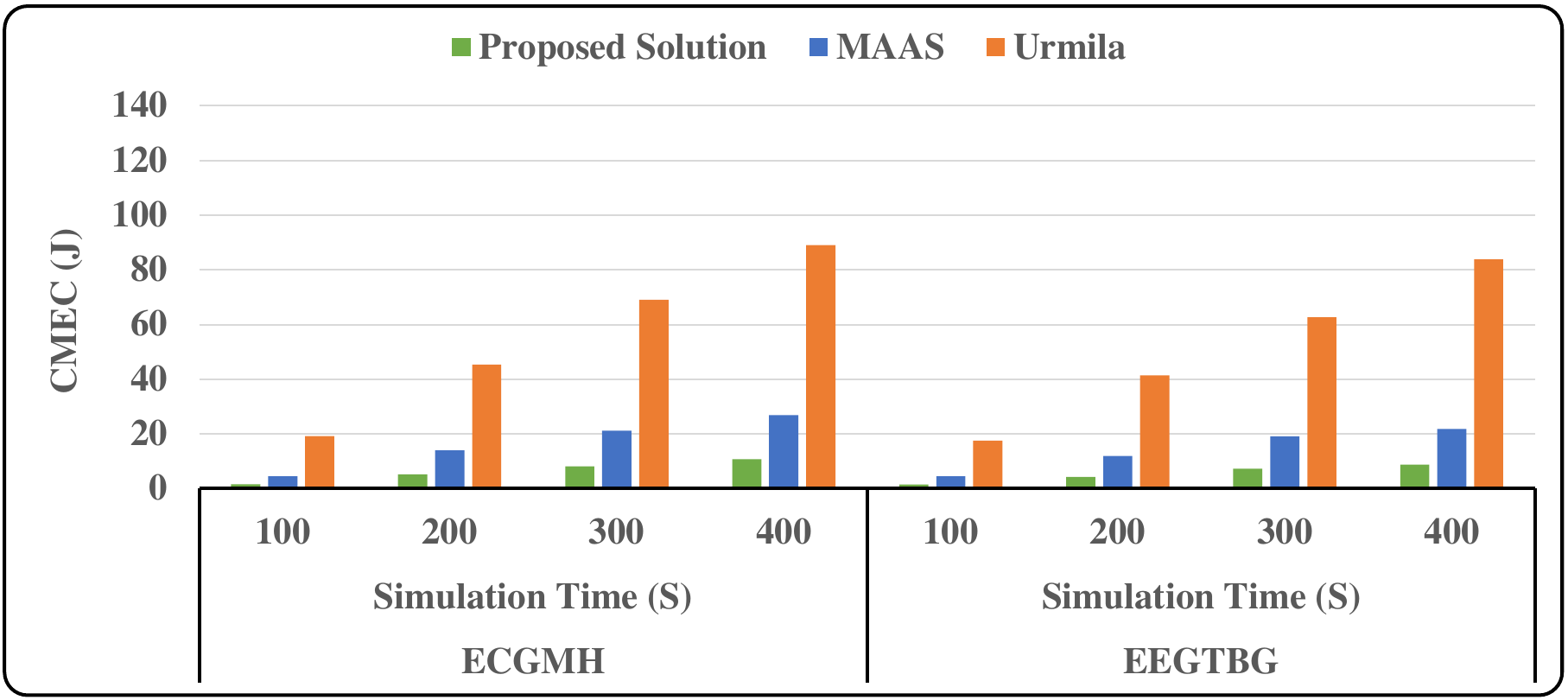}
		\captionsetup{justification=centering}
		\caption{Cumulative Migration Energy Consumption (CMEC)}
		\label{fig:CMC:sub2}
	\end{subfigure}
	\hspace{0.25mm}
	\begin{subfigure}{.325\textwidth}
		\centering
		\includegraphics[width=\linewidth,height=3cm]{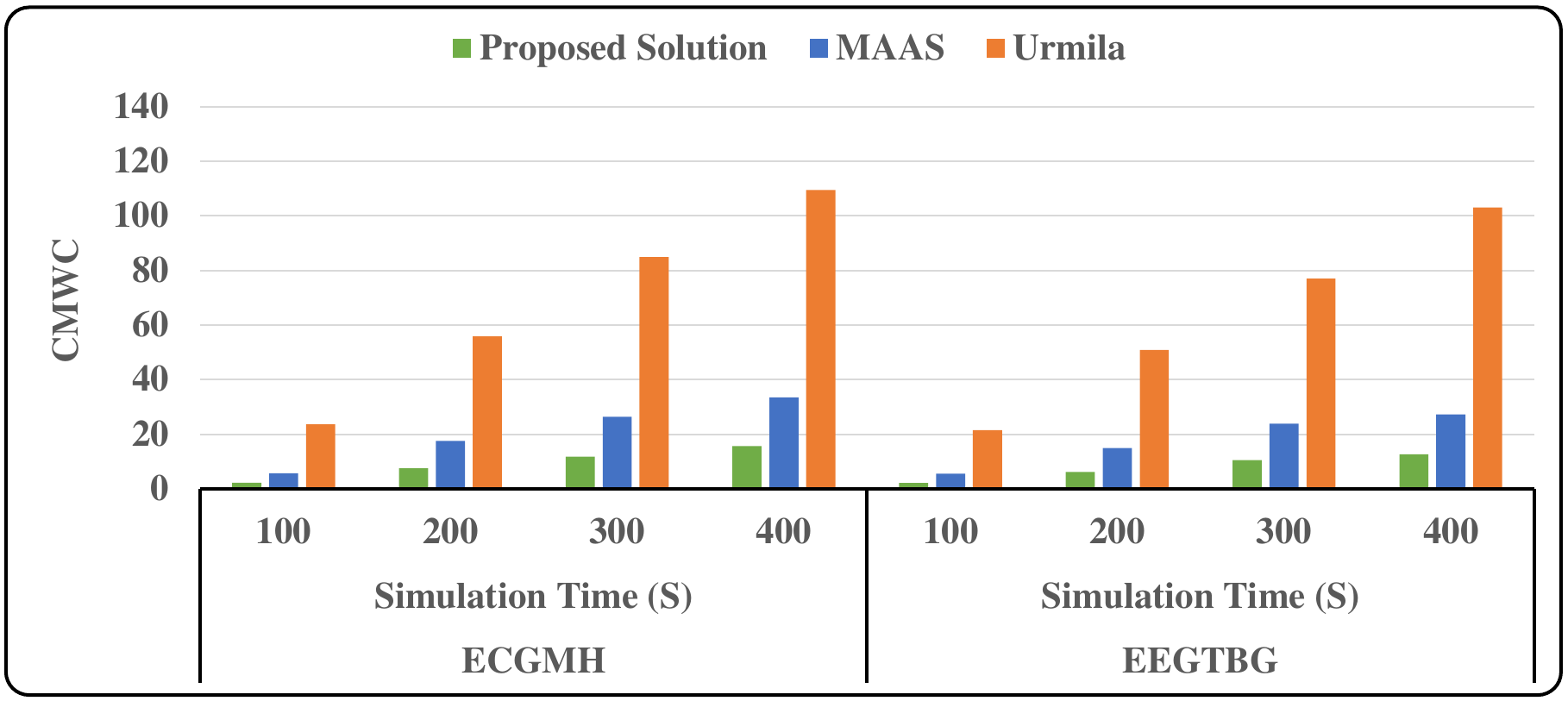}
		\captionsetup{justification=centering}
		\caption{Cumulative Migration Weighted Cost (CMWC)}
		\label{fig:CMC:sub3}
	\end{subfigure}
	
	\caption{Cumulative Migration Cost}
	\label{fig:CMC}
\end{figure*}

This experiment analyzes the Cumulative Migration Cost (CMC) of IoT devices for ECGMH and EEGTBG in different simulation times. The term cumulative refers to the aggregate migration cost of all IoT devices.
\par 
As Fig~\ref{fig:CMC} shows, our solution outperforms its counterparts in terms of Cumulative Migration Time (CMT), Cumulative Migration Energy Consumption (CMEC), and Cumulative Migration Weighted Cost (CMWC) for both ECGMH and EEGTBG applications. As the simulation time increases, the cost of all techniques grows, however, Urmila experiences a faster increase in comparison to our solution and MAAS. This latter is because the Urmila's controller is placed at the highest hierarchical layer, which incurs significant inter-nodal and transmission cost when the controller manages migrations between the old and new remote servers in the downtime. Besides, the migration cost of MAAS is more than our solution, since whenever the resources of controller finishes, the MAAS migrates the application modules to higher layers, and hence, the emitted tasks to/from those modules experience higher cost. Also, the total  number of migrations in Urmila and MAAS are higher than ours, which apparently increases their cumulative migration costs. The slight difference between cost of ECGMH and EEGTBG is because the tasks generated from the ECGMH's modules are heavier than EEGTBG's ones in terms of their MI. So, the processing time of remaining instructions of tasks (i.e., $e_{n,i,j}^{ins,r}$) that migrated from old server to new server is higher for the ECGMH compared to the EEGTBG (in case the computing powers of old and new servers are roughly the same).

\subsubsection{Total number of interrupted tasks (TIT)}
\begin{figure}[!t]
	\centering 
	\includegraphics[width=\linewidth,height=3cm]{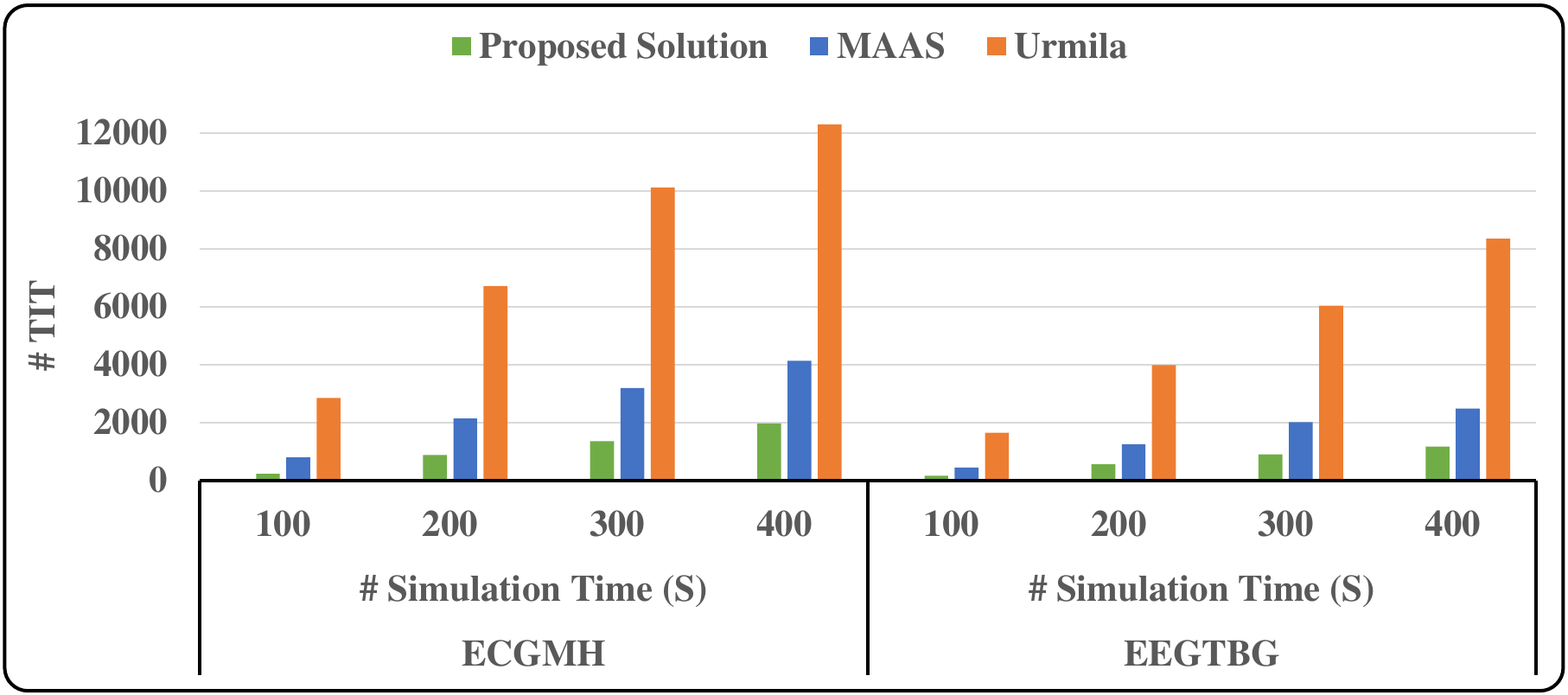}
	\caption{Total number of interrupted tasks}
	\label{fig:totalDelTask}
	
\end{figure}
This experiment analyzes the Total number of Interrupted Tasks (TIT) in the downtime. During migration downtime, there is no active service provider for incoming tasks from the modules deployed on the IoT device for a while. Hence, service interruptions happen in the downtime, in which the generated tasks experience higher delays or even they can be discarded, compared to the tasks that are generated when there is no migration. The IoT users receive smoother results with lower TIT.
\par  
Fig.~\ref{fig:totalDelTask} presents the TIT of techniques for ECGMH and EEGTBG in different simulation times. It can be seen that our solution outperforms its counterparts in different simulation times for the ECGMH and EEGTBG. The migration time has a direct impact on the TIT, and the techniques with higher migration time lead to larger TIT. This latter is because as the migration time increases, the number of delayed (or even dropped) tasks grows faster. It can be seen from Fig.~\ref{fig:totalDelTask} that the Urmila results in larger TIT than two other techniques because of its higher migration time. Moreover, due to our smaller migration time, the TIT of our solution is smaller than other techniques for both ECGMH and EEGTBG applications. It is worth mentioning that the TIT of techniques for EEGTBG applications is smaller than of ECGMH ones. This latter is due to a higher data transmission interval for the EEG sensor in EEGTBG compared to the ECG sensor of ECGMH, which means that the number of emitted tasks per second for the EEGTBG application is smaller than the ECGMH application. Hence, applications with shorter task emission interval (here, the ECGMH application) suffer more from higher migration time.

\subsubsection{Failure recovery analysis}
In this experiment, we study the effect of the failure recovery method in the migration process. The MAAS and Urmila do not have any failure recovery methods and their results are just presented here for comparison purposes. The results of our technique with a failure recovery method (FR Mode) are presented in Table~\ref{tab:Failurerecovery}  when there is a 5\% probability of failure in the migration process.
\par
Table~\ref{tab:Failurerecovery} illustrates that our technique with the failure recovery method (FR Mode) can recover from failures while it still outperforms its counterparts in terms of the total number of migrations and TIT. The obtained results of the average execution cost of tasks and cumulative migration cost in the FR Mode are roughly the same with the Non-FR Mode and they are not provided here. Since the Urmila and MAAS do not have any failure recovery methods, in case of any failures, their placement and/or migration process remains incomplete. However, in our technique, we embedded the failure recovery method for which it accepts a small overhead while it does not stop working if any failures occur.  	
\begin{table}
	\centering
	\caption{Failure Recovery Analysis}
	\label{tab:Failurerecovery}
	\resizebox{1\linewidth}{!}{%
		\renewcommand{\arraystretch}{1.2}
		\begin{tabular}{ccccc} 
			\hline
			\multirow{3}{*}{Applications} & \multirow{3}{*}{Experiment}   & \multicolumn{3}{c}{Techniques}                                                                                                                                                           \\ 
			\cline{3-5}
			&                               & \begin{tabular}[c]{@{}c@{}}Proposed Solution\\(FR Mode)~\end{tabular} & \begin{tabular}[c]{@{}c@{}}MAAS\\(No FR)\end{tabular} & \begin{tabular}[c]{@{}c@{}}Urmila\\(No FR)\end{tabular}  \\ 
			\hline
			\multirow{2}{*}{ECGMH}        & Total Number of Migrations    & 177                                                                   & 234                                                   & 234                                                      \\ 
			\cline{2-5}
			& Total Number of Interrupted Tasks & 2095                                                                  & 4152                                                  & 12302                                                    \\ 
			\hline
			\multirow{2}{*}{EEGTBG}       & Total Number of Migrations    & 169                                                                   & 227                                                   & 227                                                      \\ 
			\cline{2-5}
			& Total Number of Interrupted Tasks & 1228                                                                  & 2504                                                  & 8361                                                     \\
			\hline
		\end{tabular}
	}
\end{table}	

\subsubsection{Optimality analysis}
\begin{figure*}[!ht]
	\begin{subfigure}{.325\textwidth}
		\centering
		\includegraphics[width=\linewidth,height=3cm]{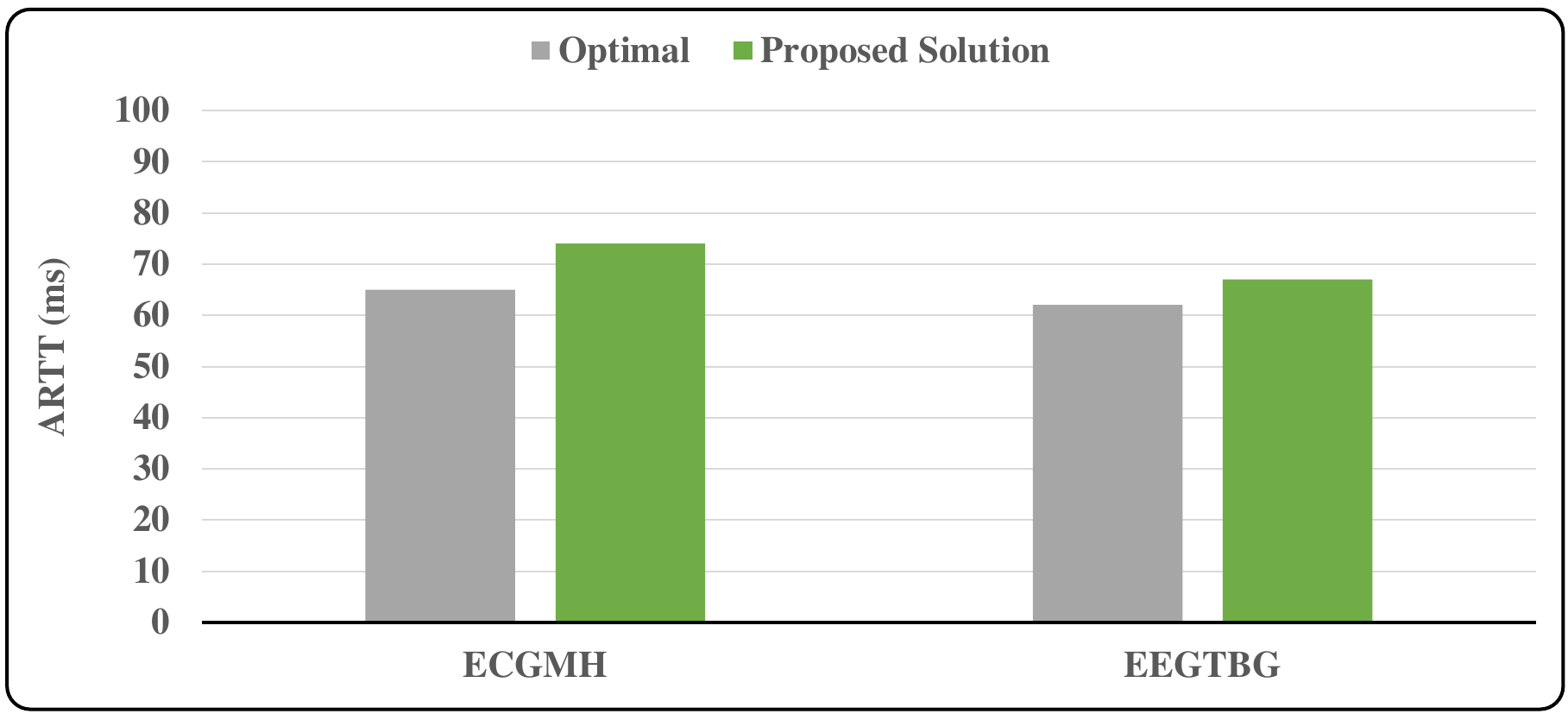}
		\captionsetup{justification=centering}
		\caption{Average Response Time of Tasks (ARTT)}
		\label{fig:OAR:sub1}
	\end{subfigure}%
	\hspace{0.1cm}
	\begin{subfigure}{.325\textwidth}
			\vspace{0.35cm}
		\centering
		\includegraphics[width=\linewidth,height=3cm]{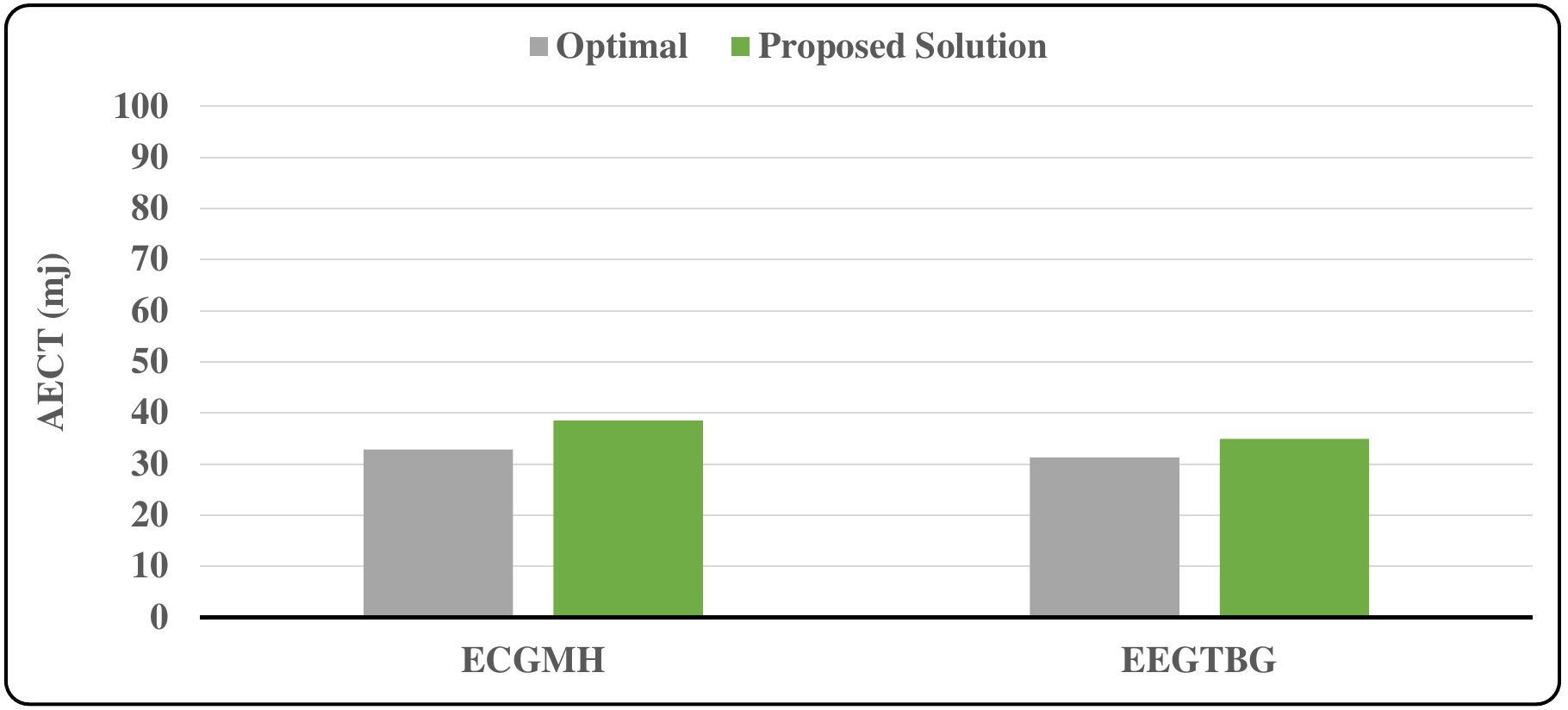}
		\captionsetup{justification=centering}
		\caption{Average Energy Consumption of Tasks (AECT)}
		\label{fig:OAR:sub2}
	\end{subfigure}
	\hspace{0.25mm}
	\begin{subfigure}{.325\textwidth}
		\centering
		\includegraphics[width=\linewidth,height=3cm]{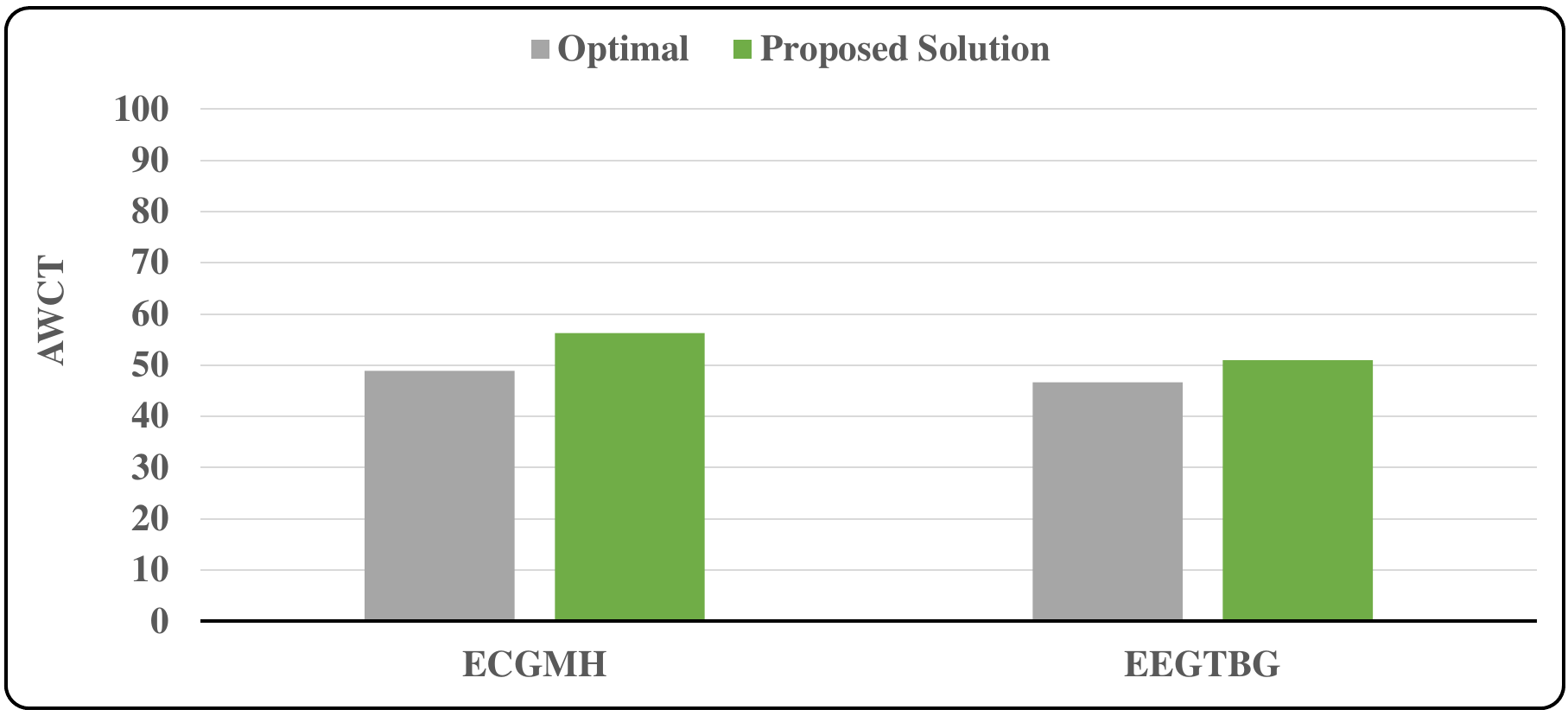}
		\captionsetup{justification=centering}
		\caption{Average Weighted Cost of Tasks (AWCT)}
		\label{fig:OAR:sub3}
	\end{subfigure}
	
	\caption{Optimality analysis results}
	\label{fig:OAR}
\end{figure*}

In this experiment, we compare the performance of our proposed solution with the optimal values. To obtain the optimal results, we used an optimized version of the Branch-and-Bound algorithm to search all possible candidate configurations for application placement, in which the bounding function helps to faster prune the search space \cite{goudarzi2017fast}. Since finding the optimal solution is very time consuming, in this experiment, we only consider 20 IoT devices in a hierarchical fog computing environment consisting of 15 candidate servers.
\par
Fig.\ref{fig:OAR} shows the results of optimality analysis in terms of Average Response Time of Tasks (ARTT), Average Energy Consumption of Tasks (AECT), and  Average Weighted Cost of Tasks (AWCT). The results show that our solution has an average of $12\%$ difference with the optimal results. However, considering the large number of FSs distributed in the proximity of IoT users, obtaining the optimal solutions, due to their large search spaces, is not practically possible, especially for real-time IoT applications. 

\section{Conclusions and Future Work}
\label{conclusion}
We proposed a new weighted cost model for minimizing the overall response time and energy consumption of IoT devices in a hierarchical fog computing environment, in which heterogeneous FSs and CSs provide services for IoT devices. In order to enable collaboration among remote servers and provide better services for IoT applications, we proposed a dynamic and distributed clustering technique among FSs of the same hierarchical level. Considering the heterogeneous resources of remote servers and the dynamic nature of such computing environments, we also proposed a distributed application placement technique to place interdependent modules of IoT applications on appropriate remote servers while satisfying their resource requirements. Also, to manage potential migrations of IoT applications' modules among remote servers, due to IoT users' mobility, a distributed migration management technique is proposed. The main goal of this latter is to reduce the migration cost of IoT applications. Finally, we embedded light-weight failure recovery methods to handle possible unpredicted failures that may happen in such dynamic computing environments. The effectiveness of our technique is analyzed through extensive experiments and comparisons by the state-of-the-art techniques in the literature. The obtained results demonstrate that our technique improves its counterparts in terms of placement deployment time, average execution cost of tasks, the total number of migrations, cumulative migration cost of all IoT devices, and the total number of interrupted tasks due to migration. 
\par  
As part of future work, we will extend our cost model to consider the energy consumption of servers and monetary cost. Moreover, we plan to consider different migration models such as pre-copy, post-copy, and hybrid, and analyze how they affect IoT applications with different resource requirements. Finally, we plan to integrate these techniques in real container-based distributed frameworks such as FogBus2 \cite{deng2021fogbus2} framework to better analyze proposed techniques in real-world scenarios.


\bibliographystyle{IEEEtran}

\bibliography{ref}


\end{document}